\documentclass[12pt]{article}
\usepackage{a4wide}
\usepackage{latexsym}
\usepackage{amsmath}
\usepackage{amsfonts}
\usepackage{amscd}
\usepackage{cite}
\usepackage{graphicx}
\usepackage{axodraw}

\usepackage{pslatex}
\usepackage[latin1]{inputenc}
\usepackage[T1]{fontenc}

\def\bq{\begin{eqnarray}}
\def\eq{\end{eqnarray}}
\def\l{\langle}
\def\r{\rangle} 
\def\eps{\varepsilon}

\begin{document}

\thispagestyle{empty}

\begin{flushright}
  MZ-TH/10-12 
\end{flushright}

\vspace{1.5cm}

\begin{center}
  {\Large\bf Introduction to Feynman Integrals\\
  }
  \vspace{1cm}
  {\large Stefan Weinzierl\\
  \vspace{6mm}
      {\small \em Institut f{\"u}r Physik, Universit{\"a}t Mainz,}\\
      {\small \em D - 55099 Mainz, Germany}\\
  } 
\end{center}

\vspace{2cm}

\begin{abstract}\noindent
  {
In these lectures I will give an introduction to Feynman integrals.
In the first part of the course I review the basics of the perturbative expansion in quantum field theories.
In the second part of the course I will discuss more advanced topics: Mathematical aspects of loop integrals
related to periods, shuffle algebras and multiple polylogarithms are covered as well as practical 
algorithms for evaluating Feynman integrals.
   }
\end{abstract}

\vspace*{\fill}

\newpage

\section{Introduction}
\label{sect:intro}

In these lectures I will give an introduction to perturbation theory and Feynman integrals
occurring in quantum field theory.
But before embarking onto a journey of integration and special function theory, it is worth
recalling the motivation for such an effort.

High-energy physics is successfully described by the Standard Model.
The term ``Standard Model'' has become a synonym for a quantum field theory
based on the gauge group $SU(3) \otimes SU(2) \otimes U(1)$.
At high energies all coupling constants are small and perturbation theory
is a valuable tool to obtain predictions from the theory.
For the Standard Model there are three coupling constants, $g_1$, $g_2$ and $g_3$, corresponding to the 
gauge groups $U(1)$, $SU(2)$ and $SU(3)$, respectively.
As all methods which will be discussed below, do not depend on the specific nature of these gauge groups
and are even applicable to extensions of the Standard Model (like super-symmetry), I will just
talk about a single expansion in a single coupling constant.
All observable quantities are taken as a power series expansion in the coupling constant, and calculated
order by order in perturbation theory.

Over the years particle physics has become a field where precision measurements have become possible.
Of course, the increase in experimental precision has to be matched with more accurate calculations
from the theoretical side.
This is the ``raison d'\^etre'' for loop calculations: A higher accuracy is reached by including more terms
in the perturbative expansion.
There is even an additional ``bonus'' we get from loop calculations:
Inside the loops we have to take into account all particles which could possibly circle there, even
the ones which are too heavy to be produced directly in an experiment.
Therefore loop calculations in combination with precision measurements allow us to extend
the range of sensitivity of experiments 
from the region which is directly accessible towards the range of heavier particles which manifest themselves
only through quantum corrections.
As an example, the mass of top quark has been predicted before the
discovery of the top quark from the loop corrections to 
electro-weak precision experiments.
The same experiments predict currently a range for the mass of the yet undiscovered Higgs boson.

It is generally believed that a perturbative series is only an asymptotic series, which will
diverge, if more and more terms beyond a certain order are included.
However this shall be of no concern to us here. We content ourselves to the first few terms in the perturbative
expansion with the implicit assumption, that the point where the power series starts to diverge is far beyond
our computational abilities.
In fact, our computational abilities are rather limited.
The complexity of a calculation increases obviously with the number of loops, but also with the number of external particles
or the number of non-zero internal masses associated to propagators.
To give an idea of the state of the art, specific quantities which are just pure numbers have been computed
up to an impressive fourth or third order.
Examples are
the calculation of the 4-loop contribution to the 
QCD $\beta$-function \cite{vanRitbergen:1997va}, 
the calculation of the anomalous magnetic moment of the electron 
up to three loops \cite{Laporta:1996mq},
and the calculation of the ratio
\bq
R & = & \frac{\sigma( e^+ e^- \rightarrow \mbox{hadrons})}
             {\sigma( e^+ e^- \rightarrow \mu^+ \mu^-)}
\eq
of the total cross section for hadron production to the total
cross section for the production of a $\mu^+ \mu^-$ pair
in electron-positron annihilation to order $O\left( g_3^3 \right)$
(also involving a three loop calculation) \cite{Gorishnii:1991vf}.
Quantities which depend on a single variable are known at the best to the third order. 
Outstanding examples are the computation
of the three-loop Altarelli-Parisi splitting functions 
\cite{Moch:2004pa,Vogt:2004mw}
or
the calculation of the two-loop amplitudes for the most interesting
$2 \rightarrow 2$ processes 
\cite{Bern:2000dn,Bern:2000ie,Bern:2001df,Bern:2001dg,Bern:2002tk,Anastasiou:2000kg,Anastasiou:2000ue,Anastasiou:2000mv,Anastasiou:2001sv,Glover:2001af,Binoth:2002xg}.
The complexity of a two-loop computation increases, if the result depends on more than one variable.
An example for a two-loop calculation whose result depends on two variables is the computation of the
two-loop amplitudes for $e^+ e^- \rightarrow \mbox{3 jets}$
\cite{Garland:2001tf,Garland:2002ak,Moch:2002hm,GehrmannDeRidder:2007bj,GehrmannDeRidder:2007hr,GehrmannDeRidder:2008ug,GehrmannDeRidder:2009dp,Weinzierl:2008iv,Weinzierl:2009ms,Weinzierl:2009nz,Weinzierl:2009yz}.

On the other hand is the mathematics encountered in these calculations of interest in its own right
and has led in the last years to a fruitful interplay between mathematicians and 
physicists.
Examples are the relation of Feynman integrals to periods, mixed Hodge structures and motives, as well as
the occurrence of certain transcendental constants in the result of a calculation\cite{Belkale:2003,Bloch:2005,Bloch:2008jk,Bloch:2008,Brown:2008,Brown:2009a,Brown:2009b,Schnetz:2008mp,Schnetz:2009,Aluffi:2008sy,Aluffi:2008rw,Aluffi:2009b,Aluffi:2009a,Bergbauer:2009yu,Laporta:2002pg,Laporta:2004rb,Laporta:2008sx,Bailey:2008ib,Bierenbaum:2003ud,Bogner:2007mn}.
Typical transcendental constants which occur in the final results are multiple zeta values. They are obtained from multiple
polylogarithms at special values of the arguments. I will discuss these functions 
in detail in these lectures.

The outline of this course is as follows:
In section~\ref{sect:action} I review the basics of perturbative quantum field theory
and I give a brief outline how Feynman rules are derived from the Lagrangian of the theory.
Issues related to the regularisation of otherwise divergent integrals 
are treated in section~\ref{sect:regularisation}.
Section~\ref{sect:loopint} is devoted to basic techniques, which allow us to exchange the integrals over the
loop momenta against integrals over Feynman parameters.
In sect.~\ref{sect:multi_loop} I discuss how the Feynman parametrisation for a generic scalar $l$-loop
integral can be read off directly from the underlying Feynman graph.
The first part of this course closes with ~\ref{sect:finite}, which shows how
finite results are obtained within perturbation theory.
The remaining section are more mathematical in nature: 
Section~\ref{sect:periods} states a general theorem which relates
Feynman integrals to periods.
Shuffle algebras and multiple polylogarithms are treated in section~\ref{sect:shuffle} 
and section~\ref{sect:polylog}, respectively.
In sect.~\ref{sect:calc} we discuss how multiple polylogarithms emerge in the calculation of Feynman
integrals.
Finally, section~\ref{sect:conclusions} provides a summary.

\section{Basics of perturbative quantum field theory}
\label{sect:action}

Elementary particle physics is described by quantum field theory. To begin with let us start with a single field
$\phi(x)$. Important concepts in quantum field theory are the Lagrangian, the action and the generating functional.
If $\phi(x)$ is a scalar field, a typical Lagrangian is
\bq
{\cal L} & = & 
 \frac{1}{2} \left( \partial_\mu \phi(x) \right) \left( \partial^\mu \phi(x) \right) - \frac{1}{2} m^2 \phi(x)^2
 + \frac{1}{4} \lambda \phi(x)^4.
\eq
The quantity $m$ is interpreted as the mass of the particle described by the field $\phi(x)$, the quantity $\lambda$
describes the strength of the interactions among the particles.
Integrating the Lagrangian over Minkowski space yields the action:
\bq
S\left[\phi \right] & = & \int d^4x \; {\cal L}\left(\phi \right).
\eq
The action is a functional of the field $\phi$. In order to arrive at the generating functional we introduce
an auxiliary field $J(x)$, called the source field, and integrate over all field configurations $\phi(x)$:
\bq
Z[J] & = & {\cal N} \int {\cal D} \phi \; e^{i \left( S[\phi] + \int d^4x J(x) \phi(x) \right)}.
\eq
The integral over all field configurations is an infinite-dimensional integral. It is called a path integral.
The prefactor ${\cal N}$ is chosen such that $Z[0]=1$.
The $n$-point Green function is given by
\bq
 \langle 0| T( \phi(x_{1}) ... \phi(x_{n})) |0 \rangle & = &
   \frac{\int {\cal D} \phi \; \phi(x_{1}) ... \phi(x_{n}) e^{i S(\phi)}}
        {\int {\cal D} \phi \; e^{i S(\phi)}}.
\eq
With the help of functional derivatives this can be expressed as
\bq
 \langle 0| T( \phi(x_{1}) ... \phi(x_{n})) |0 \rangle & = &
   \left. \left(-i \right)^n
   \frac{\delta^n Z[J]}{\delta J(x_1) ... \delta J(x_n)} \right|_{J=0}.
\eq
We are in particular interested in connected Green functions. These are obtained from a functional $W[J]$, which is
related to $Z[J]$ by
\bq
Z[J] & = & e^{i W[J]}.
\eq
The connected Green functions are then given by
\bq
\label{functional_derivative}
G_n(x_1,...,x_n) & = & \left( -i \right)^{n-1} 
\left. \frac{\delta^n W[J]}{\delta J(x_1) ... \delta J(x_n) } \right|_{J=0}.
\eq
It is convenient to go from position space to momentum space by a Fourier transformation.
We define the Green functions in momentum space by
\bq
\label{fourier_trafo}
G_n(x_1,...,x_n) & = & \int \frac{d^4p_1}{(2\pi)^4} ... \frac{d^4p_n}{(2\pi)^4}
 e^{-i \sum p_j x_j} \left(2 \pi \right)^4 \delta\left(p_1+...+p_n\right) \tilde{G}_n(p_1,...,p_n).
\eq
Note that the Fourier transform $\tilde{G}_n$ is defined by explicitly factoring out the 
$\delta$-function $\delta(p_1+...+p_n)$ and a factor $(2 \pi )^4$.
We denote the two-point function in momentum space by $\tilde{G}_2(p)$.
In this case we have to specify only one momentum, since the momentum flowing into the Green function on one side
has to be equal to the momentum flowing
out of the Green function on the other side due to the presence of the $\delta$-function in eq.~(\ref{fourier_trafo}) .
We now are in a position to define the scattering amplitude: In momentum space the scattering amplitude 
with $n$ external particles is
given by the connected $n$-point Green function multiplied by the inverse two-point function for each external
particle:
\bq
{\cal A}_n\left(p_1,...,p_n\right)
 & = &
 \tilde{G}_2\left(p_1\right)^{-1}
 ...
 \tilde{G}_2\left(p_n\right)^{-1}
 \tilde{G}_n(p_1,...,p_n).
\eq
The scattering amplitude enters directly the calculation of a physical observable.
Let us first consider the scattering process of two incoming particles with four-momenta $p_1'$ and $p_2'$ 
and $(n-2)$ outgoing particles with four-momenta $p_1$ to $p_{n-2}$.
Let us assume that we are interested in an observable $O\left(p_1,...,p_{n-2}\right)$ 
which depends on the momenta of the outgoing particles.
In general the observable
depends on the experimental set-up and 
can be an arbitrary complicated function of the four-momenta. 
In the simplest case this function is just a constant equal to one,
corresponding to the situation where we count every event with $(n-2)$ particles in the final state.
In more realistic situations one takes for example into account that it is not possible to detect 
particles close to the beam pipe. The function $O$ would then be zero in these regions of phase space.
Furthermore any experiment has a finite resolution. Therefore it will not be possible to detect particles
which are very soft or which are very close in angle to other particles.
We will therefore sum over the number of final state particles.
In order to obtain finite results within perturbation theory we have to require that 
in the case where one or more particles become unresolved 
the value of the observable ${\cal O}$ has a continuous limit
agreeing with the value of the observable for a configuration where the unresolved particles have been merged 
into ``hard'' (or resolved) particles.
Observables having this property are called infrared-safe observables. 
The expectation value for the observable $O$ is given by
\bq
\label{observable_master}
\langle O \rangle & = & \frac{1}{2 (p_1'+p_2')^2}
 \sum\limits_n
             \int d\phi_{n-2}
             O\left(p_1,...,p_{n-2}\right)
             \left| {\mathcal A}_n \right|^2,
\eq
where $1/2/(p_1'+p_2')^2$ is a normalisation factor taking into account the incoming flux.
The phase space measure is given by
\bq
\label{phasespacemeasure}
d\phi_n & = &
 \frac{1}{n! }
 \prod\limits_{i=1}^n \frac{d^{3}p_i}{(2 \pi)^{3} 2 E_i} 
 \left(2 \pi \right)^4 \delta^4\left(p_1'+p_2'-\sum\limits_{i=1}^n p_i\right).
\;\;\;\;\;\;\;
\eq
$E_i$ is the energy of particle $i$:
\bq
E_i & = & \sqrt{\vec{p}_i^2 +m_i^2}
\eq
We see that the expectation value of $O$ is given by the phase space integral over the observable, weighted
by the norm squared of the scattering amplitude.
As the integrand can be a rather complicated function, the phase space integral is usually performed numerically by
Monte Carlo integration.

Let us now look towards a more realistic theory. 
As an example I will take quantum chromodynamics (QCD), which describes the strong force and which is
formulated in terms of quarks and gluons. Quarks and gluons are collectively called partons.
There are a few modifications to eq.~(\ref{observable_master}). The master formula reads now
\bq
\label{observable_master_hadron}
\lefteqn{
\l O \r = 
 \sum\limits_{a,b}
 \int dx_1 f_a(x_1) \int dx_2 f_b(x_2) 
} & & 
 \\
 & & 
             \frac{1}{2 \hat{s} n_s(1) n_s(2) n_c(1) n_c(2)}
 \sum\limits_n
             \int d\phi_{n-2}
             O\left(p_1,...,p_{n-2}\right)
             \sum\limits_{\mathrm{spins,colour}} 
             \left| {\cal A}_n \right|^2.
 \nonumber
\eq
The partons have internal degrees of freedom, given by the spin and the colour of the partons.
In squaring the amplitude we sum over these degrees of freedom. 
For the particles in the initial state we would like to average over these degrees of freedom.
This is done by dividing by the factors $n_s(i)$ and $n_c(i)$, giving the number of spin degrees of
freedom ($2$ for quarks and gluons)
and the number of colour degrees of freedom ($3$ for quarks, $8$ for gluons).
The second modification is due to the fact that the particles brought into collision are not partons, but 
composite particles like protons. At high energies the constituents of the protons interact and we have to include
a function $f_a(x)$ giving us the probability of finding a parton $a$ with momentum fraction $x$ of the original
proton momentum inside the proton.
$\hat{s}$ is the centre-of-mass energy squared of the two partons entering the hard interaction.
In addition there is a small change in eq.~(\ref{phasespacemeasure}). The quantity $(n!)$ is replaced by
$(\prod n_j!)$, where $n_j$ is the number of times a parton of type $j$ occurs in the final state.

As before, the scattering amplitude ${\cal A}_n$ can be calculated once the Lagrangian of the theory 
has been specified. For QCD the Lagrange density reads:
\bq
{\cal L}_{\mathrm{QCD}} & = & 
 - \frac{1}{4} F^{a}_{\mu\nu}(x) F^{a \mu\nu}(x)
 - \frac{1}{2 \xi} ( \partial^{\mu} A^{a}_{\mu}(x) )^{2} 
 +
 \sum\limits_{\mathrm{quarks } \; q} 
   \bar{\psi}_q(x) \left(  i \gamma^{\mu} D_{\mu} - m_q \right) \psi_q(x)
 + {\cal L}_{\mathrm{FP}}, 
 \;\;\;\;\;
\eq
with
\bq
F^{a}_{\mu\nu}(x) = \partial_{\mu} A^{a}_{\nu}(x) - \partial_{\nu} A^{a}_{\mu}(x)
 + g f^{abc} A^{b}_{\mu}(x) A^{c}_{\nu},
 & &
 D_\mu = \partial_\mu - i g T^a A^a_\mu(x).
\eq
The gluon field is denoted by $A_\mu^a(x)$, the quark fields are denoted by $\psi_q(x)$.
The sum is over all quark flavours. The masses of the quarks are denoted by $m_q$.
There is a summation over the colour indices of the quarks, which is not shown explicitly.
The variable $g$ gives the strength of the strong coupling.
The generators of the group $SU(3)$ are denoted by $T^a$ and satisfy
\bq
 \left[ T^a, T^b \right] & = & i f^{abc} T^c.
\eq
The quantity $F^a_{\mu\nu}$ is called the field strength, the quantity $D_\mu$ is called the covariant derivative.
The variable $\xi$ is called the gauge-fixing parameter. Gauge-invariant quantities like scattering amplitudes are
independent of this parameter.
${\cal L}_{\mathrm{FP}}$ stands for the Faddeev-Popov term, which arises through the gauge-fixing procedure and which is only
relevant for loop amplitudes.

Unfortunately it is not possible to calculate from this Lagrangian the scattering amplitude ${\cal A}_n$
exactly. The best what can be done is to expand the scattering amplitude in the small parameter $g$ and to calculate
the first few terms.
The amplitude ${\cal A}_n$ with $n$ external partons has the perturbative expansion
\bq
\label{basic_perturbative_expansion}
 {\mathcal A}_n & = & g^{n-2} \left( {\mathcal A}_n^{(0)} + g^2 {\mathcal A}_n^{(1)} + g^4 {\mathcal A}_n^{(2)} + g^6 {\mathcal A}_n^{(3)} + ... \right).
\eq
In principle we could now calculate every term in this expansion by taking the functional derivatives according to
eq.~(\ref{functional_derivative}).
This is rather tedious and there is a short-cut to arrive at the same result, which is based on Feynman graphs.
The recipe for the computation of ${\mathcal A}_n^{(l)}$ is as follows: Draw first all Feynman diagrams with the given number
of external particles and $l$ loops. Then translate each graph into a mathematical formula with the help of the Feynman
rules.
${\mathcal A}_n^{(l)}$ is then given as the sum of all these terms.

In order to derive the Feynman rules from the Lagrangian one proceeds as follows:
One first separates the Lagrangian into a part which is bilinear in the fields, and a part where each
term contains three or more fields.
(A ``normal'' Lagrangian does not have parts with just one or zero fields.)
From the part bilinear in the fields one derives the propagators, while the terms with three or more
fields give rise to vertices.
As an example we consider the gluonic part of the QCD Lagrange density:
\bq
\label{Lagrangian_QCD_expanded}
{\cal L}_{\mathrm{QCD}} & = & 
 \frac{1}{2} A^{a}_{\mu}(x) \left[ \partial_\rho \partial^\rho g^{\mu\nu} \delta^{ab}
                  - \left( 1 - \frac{1}{\xi} \right) \partial^\mu \partial^\nu \delta^{ab} \right] A^{b}_{\nu}(x)
 \nonumber \\
 & &
 - g f^{abc} \left( \partial_\mu A^{a}_{\nu}(x) \right) A^{b \mu}(x) A^{c \nu}(x)
 - \frac{1}{4} g^2 f^{eab} f^{ecd} A^{a}_{\mu}(x) A^{b}_{\nu}(x) A^{c \mu}(x) A^{d \nu}(x) 
 \nonumber \\
 & &
 + {\cal L}_{\mathrm{quarks}}
 + {\cal L}_{\mathrm{FP}}.
\eq
Within perturbation theory we always assume that all fields fall off rapidly enough at infinity.
Therefore we can ignore boundary terms within partial integrations.
The expression in the first line is bilinear in the fields. The terms in the square bracket in this line
define an operator
\bq
 P^{\mu\nu\;ab}(x) & = & \partial_\rho \partial^\rho g^{\mu\nu} \delta^{ab}
                  - \left( 1 - \frac{1}{\xi} \right) \partial^\mu \partial^\nu \delta^{ab}.
\eq
For the propagator we are interested in the inverse of this operator
\bq
 P^{\mu\sigma\;ac}(x) \left( P^{-1} \right)_{\sigma\nu}^{cb}(x-y) & = & g^\mu_{\;\;\nu} \delta^{ab} \delta^4(x-y).
\eq
Working in momentum space we are more specifically interested in the Fourier transform of the inverse
of this operator:
\bq
 \left( P^{-1} \right)_{\mu\nu}^{ab}(x) & = & 
  \int \frac{d^4 k}{(2 \pi)^4} e^{-i k \cdot x} \left( \tilde{P}^{-1} \right)_{\mu\nu}^{ab}(k).
\eq
The Feynman rule for the propagator is then given by $(\tilde{P}^{-1})_{\mu\nu}^{ab}(k)$ times the imaginary unit.
For the gluon propagator one finds the Feynman rule
\bq
 \begin{picture}(100,20)(0,5)
 \Gluon(20,10)(70,10){-5}{5}
 \Text(15,12)[r]{\footnotesize $\mu, a$}
 \Text(75,12)[l]{\footnotesize $\nu, b$}
\end{picture} 
 & = & 
  \frac{i}{k^2} \left( - g_{\mu\nu} + \left( 1 -\xi \right) \frac{k_\mu k_\nu}{k^2} \right) \delta^{ab}.
\eq
To derive the Feynman rules for the vertices we look as an example at the first term in the second line of
eq.~(\ref{Lagrangian_QCD_expanded}):
\bq
 {\cal L}_{ggg} & = &
 - g f^{abc} \left( \partial_\mu A^{a}_{\nu}(x) \right) A^{b \mu}(x) A^{c \nu}(x).
\eq
This term contains three gluon fields and will give rise to the three-gluon vertex.
We rewrite this term as follows:
\bq
 {\cal L}_{ggg} & = &
 \int d^4x_1 d^4x_2 d^4x_3
 \alpha^{abc\;\mu\nu\lambda}(x,x_1,x_2,x_3)
 A^{a}_\mu(x_1) A^{b}_\nu(x_2) A^{c}_\lambda(x_3),
\eq
where
\bq
 \alpha^{abc\;\mu\nu\lambda}(x,x_1,x_2,x_3) & = &
    g f^{abc} g^{\mu\lambda} \left( \partial^\nu_{x_1} \delta^4(x-x_1) \right) \delta^4(x-x_2) \delta^4(x-x_3).
\eq
Again we are interested in the Fourier transform of this expression:
\bq
 \alpha^{abc\;\mu\nu\lambda}(x,x_1,x_2,x_3) & = &
 \int \frac{d^4k_1}{(2 \pi)^4} \frac{d^4k_2}{(2 \pi)^4} \frac{d^4k_3}{(2 \pi)^4}
 e^{-i k_1 (x_1-x) -i k_2 (x_2-x) - i k_3 (x_3-x)} 
 \tilde{\alpha}^{abc\;\mu\nu\lambda}(k_1,k_2,k_3).
 \nonumber
\eq
Working this out we find
\bq
 \tilde{\alpha}^{abc\;\mu\nu\lambda}(k_1,k_2,k_3)
 & = &
    - g f^{abc} g^{\mu\lambda} i k_1^\nu.
\eq
The Feynman rule for the vertex is then given by the sum over all permutations of identical particles of the
function $\tilde{\alpha}$ multiplied by the imaginary unit $i$. 
(In the case of identical fermions there would be in addition a minus sign for every odd permutation 
of the fermions.)
We thus obtain the Feynman rule for the three-gluon vertex:
\bq
\begin{picture}(100,35)(0,55)
\Vertex(50,50){2}
\Gluon(50,50)(50,80){3}{4}
\Gluon(50,50)(76,35){3}{4}
\Gluon(50,50)(24,35){3}{4}
\LongArrow(56,70)(56,80)
\LongArrow(67,47)(76,42)
\LongArrow(33,47)(24,42)
\Text(60,80)[lt]{$k_1^\mu,a$}
\Text(78,35)[lc]{$k_2^\nu,b$}
\Text(22,35)[rc]{$k_3^\lambda,c$}
\end{picture}
 & = &
g f^{abc} \left[ 
                 g^{\mu \nu} \left(k_2^\lambda - k_1^\lambda \right) 
               + g^{\nu \lambda} \left(k_3^\mu - k_2^\mu \right)  
               + g^{\lambda \mu} \left(k_1^\nu - k_3^\nu \right)  
          \right].
 \\ \nonumber
 \\ \nonumber
\eq
Note that there is momentum conservation at each vertex, for the three-gluon vertex this implies
\bq
 k_1 + k_2 + k_3 & = & 0.
\eq
Following the procedures outlined above we can derive the Feynman rules for all propagators and vertices of the 
theory.
If an external particle carries spin, we have to associate a factor, which describes 
the polarisation of the corresponding particle when we translate a Feynman diagram into a formula.
Thus, there is a polarisation vector $\eps^\mu(p)$ for each external gauge boson and a spinor 
$\bar{u}(p)$, $u(p)$, $\bar{v}(p)$ or $v(p)$ for each external fermion.

Furthermore there are a few additional rules: First of all, there is an
integration
\bq
 \int \frac{d^4k}{(2\pi)^4}
\eq
for each internal momentum which is not constrained by momentum conservation.
Such an integration is called a ``loop integration'' and the number of independent loop integrations in a diagram
is called the loop number of the diagram.
Secondly, each closed fermion loop gets an extra factor of $(-1)$.
Finally, each diagram gets multiplied by a symmetry factor $1/S$,
where $S$ is the order of the permutation group
of the internal lines and vertices leaving the diagram unchanged when the external lines are fixed.

Let us finish this section by listing the remaining Feynman rules for QCD. The quark and the ghost 
propagators are given by
\bq
\begin{picture}(85,20)(0,5)
 \ArrowLine(70,10)(20,10)
 \Text(15,10)[rb]{\footnotesize $j$}
 \Text(75,10)[lb]{\footnotesize $l$}
\end{picture} 
 & = &
 i \frac{k\!\!\!/+m}{k^2-m^2} \delta_{jl},
 \nonumber \\
\begin{picture}(85,20)(0,5)
 \DashArrowLine(70,10)(20,10){3}
 \Text(15,10)[rb]{\footnotesize $a$}
 \Text(75,10)[lb]{\footnotesize $b$}
\end{picture} 
 & = &
 \frac{i}{k^2} \delta^{ab}.
\eq
The Feynman rules for the four-gluon vertex, the quark-gluon vertex and the ghost-gluon vertex are
\bq
\begin{picture}(100,35)(0,55)
\Vertex(50,50){2}
\Gluon(50,50)(71,71){3}{4}
\Gluon(50,50)(71,29){3}{4}
\Gluon(50,50)(29,29){3}{4}
\Gluon(50,50)(29,71){3}{4}
\Text(72,72)[lb]{\small $\mu,a$}
\Text(72,28)[lt]{\small $\nu,b$}
\Text(28,28)[rt]{\small $\lambda,c$}
\Text(28,72)[rb]{\small $\rho,d$}
\end{picture}
 & = &
- i g^{2} \left[ f^{abe} f^{ecd} \left( g^{\mu \lambda} g^{\nu \rho} - g^{\mu \rho} g^{\nu \lambda} 
                                 \right) 
               + f^{ace} f^{ebd} \left( g^{\mu \nu} g^{\lambda \rho} - g^{\mu \rho} g^{\lambda \nu} \right) 
 \right. \nonumber \\
 & & \left.
         + f^{ade} f^{ebc} \left( g^{\mu \nu} g^{\lambda \rho} - g^{\mu \lambda} g^{\nu \rho} \right)
\right],
 \nonumber \\ 
\begin{picture}(100,35)(0,55)
\Vertex(50,50){2}
\Gluon(50,50)(80,50){3}{4}
\ArrowLine(50,50)(29,71)
\ArrowLine(29,29)(50,50)
\Text(82,50)[lc]{$\mu$,$a$}
\Text(28,33)[rt]{$l$}
\Text(28,67)[rb]{$j$}
\end{picture}
 & = &
i g \gamma^\mu T^a_{jl},
 \nonumber \\
 \nonumber \\
 \nonumber \\
\begin{picture}(100,35)(0,55)
\Vertex(50,50){2}
\Gluon(50,50)(80,50){3}{4}
\DashArrowLine(50,50)(29,71){3}
\DashArrowLine(29,29)(50,50){3}
\LongArrow(36,59)(29,66)
\Text(82,50)[lc]{$\mu$,$b$}
\Text(28,29)[rt]{$q$,$c$}
\Text(28,71)[rb]{$k$,$a$}
\end{picture}
 & = &
 -g f^{abc} k_{\mu}.
 \\
 \nonumber \\ \nonumber
\eq
The Feynman rules for the electro-weak sector of the Standard Model are similar, but too numerous to list them 
explicitly here.

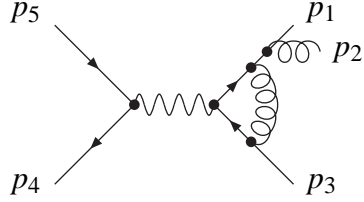
\begin{figure}
\begin{center}
\begin{picture}(100,60)(0,30)
\Photon(30,50)(60,50){4}{4}
\Vertex(60,50){2}
\ArrowLine(60,50)(75,65)
\Line(75,65)(90,80)
\Line(90,20)(75,35)
\ArrowLine(75,35)(60,50)
\GlueArc(60,50)(20,-45,45){4}{4}
\Vertex(74,64){2}
\Vertex(74,36){2}
\Vertex(80,70){2}
\Gluon(80,70)(100,70){4}{2}
\Text(95,85)[l]{$p_1$}
\Text(105,70)[l]{$p_2$}
\Text(95,20)[l]{$p_3$}
\Vertex(30,50){2}
\ArrowLine(0,80)(30,50)
\ArrowLine(30,50)(0,20)
\Text(-5,20)[r]{$p_4$}
\Text(-5,85)[r]{$p_5$}
\end{picture} 
\end{center}
\caption{\label{fig1} A one-loop Feynman diagram contributing to the process
$e^+ e^- \rightarrow q g \bar{q}$.}
\end{figure}

Having stated the Feynman rules, let us look at some examples.
We have seen that for a given process with a specified set of external particles
the scattering amplitude is given as the sum of all Feynman diagrams with this set of external particles.
We can order the diagrams by the powers of the coupling factors. In QCD we obtain for each three-particle vertex 
one power of $g$, while the four-gluon vertex contributes two powers of $g$.
The leading order result for the scattering amplitude is obtained by taking only the diagrams with the minimal number
of coupling factors $g$ into account.
These are diagrams which have no closed loops. There are no conceptual difficulties in evaluating these diagrams.
However going beyond the leading order in perturbation theory, loop diagrams appear which involve 
integrations over the loop momenta.
These diagrams are more difficult to evaluate and I will discuss them in more detail.
Fig. \ref{fig1} shows a Feynman diagram contributing to the one-loop corrections
for the process $e^+ e^- \rightarrow q g \bar{q}$.
At high energies we can ignore the masses of the electron and the light quarks.
From the Feynman rules one obtains for this diagram:
\bq
\label{feynmanrules}
- e^2 g^3 
  C_F T^a_{jl} 
  \bar{v}(p_4) \gamma^\mu u(p_5)
  \frac{1}{p_{123}^2}
  \int \frac{d^{4}k_1}{(2\pi)^{4}}
  \frac{1}{k_2^2}
  \bar{u}(p_1) \eps\!\!\!/(p_2) \frac{p\!\!\!/_{12}}{p_{12}^2}
  \gamma_\nu \frac{k\!\!\!/_1}{k_1^2}
  \gamma_\mu \frac{k\!\!\!/_3}{k_3^2}
  \gamma^\nu
  v(p_3).
\eq
Here, $p_{12}=p_1+p_2$, $p_{123}=p_1+p_2+p_3$, $k_2=k_1-p_{12}$, $k_3=k_2-p_3$.
Further $\eps\!\!\!/(p_2) = \gamma_\tau \eps^\tau(p_2)$, where $\eps^\tau(p_2)$ is the
polarisation vector of the outgoing gluon.
All external momenta are assumed to be
massless: $p_i^2=0$ for $i=1..5$.
We can reorganise this formula into a part, which depends on the loop integration and a part, which does not.
The loop integral to be calculated reads:
\bq
\label{loop_int_example_1}
  \int \frac{d^4 k_1}{(2\pi)^{4}}
  \frac{k_1^\rho k_3^\sigma}{k_1^2 k_2^2 k_3^2},
\eq
while the remainder, which is independent of the loop integration is given by
\bq
\label{loop_int_example_remainder}
- e^2 g^3 
  C_F T^a_{jl}
  \bar{v}(p_4) \gamma^\mu u(p_5)
  \frac{1}{p_{123}^2 p_{12}^2}
  \bar{u}(p_1) \eps\!\!\!/(p_2) p\!\!\!/_{12}
  \gamma_\nu \gamma_\rho
  \gamma_\mu \gamma_\sigma
  \gamma^\nu
  v(p_3).
\eq
The loop integral in eq.~(\ref{loop_int_example_1}) contains in the denominator three propagator factors
and in the numerator two factors of the loop momentum.
We call a loop integral, in which the loop momentum occurs also in the numerator a ``tensor integral''.
A loop integral, in which the numerator is independent of the loop momentum is called a ``scalar integral''.
The scalar integral associated to eq.~(\ref{loop_int_example_1}) reads
\bq
\label{loop_int_example_1a}
  \int \frac{d^4 k_1}{(2\pi)^{4}}
  \frac{1}{k_1^2 k_2^2 k_3^2}.
\eq
It is always possible to reduce tensor integrals to scalar integrals \cite{Tarasov:1996br,Tarasov:1997kx}. 
The calculation of integrals like the one in eq.~(\ref{loop_int_example_1a}) is the main topic of these lectures.
More information on the basics of perturbation theory and quantum field theory 
can be found in one of the many textbooks on quantum field theory, like
for example in refs.~\cite{Peskin,Itzykson:1980rh}.

\section{Dimensional regularisation}
\label{sect:regularisation}

Before we start with the actual calculation of loop integrals, I should mention one 
complication: Loop integrals are often divergent !
Let us first look at the simple example of a scalar two-point one-loop integral with zero 
external momentum: 
\bq
\begin{picture}(100,40)(0,30)
 \Line(10,35)(25,35)
 \Line(75,35)(90,35)
 \CArc(50,35)(25,0,360)
 \Vertex(25,35){2}
 \Vertex(75,35){2}
 \Text(5,35)[r]{$p=0$}
 \Text(50,55)[t]{\scriptsize $k$}
 \Text(50,15)[b]{\scriptsize $k$}
\end{picture}
 & = &
\int \frac{d^4k}{(2\pi)^4} \frac{1}{(k^2)^2} 
 =
\frac{1}{(4\pi)^2} \int\limits_0^\infty dk^2 \frac{1}{k^2} = 
\frac{1}{(4\pi)^2} \int\limits_0^\infty \frac{dx}{x}.
\eq
This integral diverges at $k^2\rightarrow \infty$ as well as at $k^2\rightarrow 0$.
The former divergence is called ultraviolet divergence, the later is called infrared divergence.
Any quantity, which is given by a divergent integral, is of course an ill-defined quantity.
Therefore the first step is to make these integrals well-defined by introducing a regulator.
There are several possibilities how this can be done, but the
method of dimensional regularisation 
\cite{'tHooft:1972fi,Bollini:1972ui,Cicuta:1972jf}
has almost become a standard, as the calculations in this regularisation
scheme turn out to be the simplest.
Within dimensional regularisation one replaces the four-dimensional integral over the loop momentum by an
$D$-dimensional integral, where $D$ is now an additional parameter, which can be a non-integer or
even a complex number.
We consider the result of the integration as a function of $D$ and we are interested in the behaviour of this 
function as $D$ approaches $4$.
The $D$-dimensional integration still fulfils the standard laws for integration,
like linearity, translation invariance and scaling behaviour
\cite{Wilson:1972cf,Collins}.
If $f$ and $g$ are two functions, and if $a$ and $b$ are two constants, 
linearity states that
\bq
\int d^{D} k \left( a f(k) + b g(k) \right) & = & a \int d^{D}k f(k) + b \int d^{D}k g(k).
\eq
Translation invariance requires that 
\bq
\int d^{D}k f(k+p) & = & \int d^{D}k f(k).
\eq
for any vector $p$.
\\
The scaling law states that
\bq
\int d^{D}k f(\lambda k) & = & \lambda^{-D} \int d^{D}k f(k).
\eq
The $D$-dimensional integral has also a rotation invariance:
\bq
\int d^{D}k f( \Lambda k) & = & \int d^{D}k f(k),
\eq
where $\Lambda$ is an element of the Lorentz group $SO(1,D-1)$ of the $D$-dimensional vector-space.
Here we assumed that the $D$-dimensional vector-space has the metric $\mbox{diag}(+1,-1,-1,-1,...)$.
The integral measure is normalised such that it agrees with the result for the integration of a Gaussian
function for all integer values $D$:
\bq
\label{normalisation_D_int}
\int d^{D}k \exp \left( \alpha k^2 \right) & = & 
 i \left( \frac{\pi}{\alpha} \right)^{\frac{D}{2}}.
\eq
We will further assume that we can always decompose any vector into a $4$-dimensional part and
a $(D-4)$-dimensional part
\bq
 k^\mu_{(D)} & = & k^\mu_{(4)} + k^\mu_{(D-4)},
\eq
and that the $4$-dimensional and $(D-4)$-dimensional subspaces are orthogonal to each other:
\bq
 k_{(4)} \cdot k_{(D-4)} & = & 0.
\eq
If $D$ is an integer greater than $4$, this is obvious. We postulate that these relations are true
for any value of $D$. One can think of the underlying vector-space as a space of infinite dimension, where 
the integral measure mimics the one in $D$ dimensions.

In practise we will always arrange things such that every function we integrate over $D$ dimensions
is rotational invariant, e.g. is a function of $k^2$.
In this case the integration over the $(D-1)$ angles is trivial and can be expressed in a closed form
as a function of $D$.
Let us assume that we have an integral, which has a UV-divergence, but no IR-divergences. 
Let us further assume that this integral would diverge logarithmically, if we would use
a cut-off regularisation instead of dimensional regularisation. 
It turns out that this integral will be convergent if the real part of $D$ is smaller than $4$.
Therefore we may compute this integral under the assumption that $\mbox{Re}(D)<4$ and we will obtain as
a result a function of $D$. This function can be analytically continued to the whole complex plane.
We are mainly interested in what happens close to the point $D=4$. For an ultraviolet divergent one-loop
integral we will find that the analytically continued result will exhibit a pole at $D=4$.
It should be mentioned that there are also integrals which are quadratically divergent, if a cut-off regulator
is used.
In this case we can repeat the argumentation above with the replacement $\mbox{Re}(D)<2$.

Similarly, we can consider an IR-divergent integral, which has no UV-divergence. This integral
will be convergent if $\mbox{Re}(D)>4$. Again, we can compute the integral in this domain and
continue the result to $D=4$. Here we find that each IR-divergent loop integral can lead to a double
pole at $D=4$.

We will use dimensional regularisation to regulate both the ultraviolet and infrared divergences.
The attentative reader may ask how this goes together, as we argued above that UV-divergences require
$\mbox{Re}(D)<4$ or even $\mbox{Re}(D)<2$, whereas IR-divergences are regulated by $\mbox{Re}(D)>4$.
Suppose for the moment that we use dimensional regularisation just for the UV-divergences
and that we use a second regulator for the IR-divergences.
For the IR-divergences we could keep all external momenta off-shell, or introduce small masses for all massless
particles or even raise the original propagators to some power $\nu$.
The exact implementation of this regulator is not important, as long as the IR-divergences are screened by this
procedure. We then perform the loop integration in the domain where the integral is UV-convergent.
We obtain a result, which we can analytically continue to the whole complex $D$-plane, in particular
to $\mbox{Re}(D)>4$. There we can remove the additional regulator and the IR-divergences are now regulated
by dimensional regularisation. Then the infrared divergences
will also show up as poles at $D=4$.

There is one more item which needs to be discussed in the context of dimensional regularisation:
Let us look again at the example in eqs.~(\ref{feynmanrules}) to (\ref{loop_int_example_remainder}).
We separated the loop integral from the remainder in eq.~(\ref{loop_int_example_remainder}), 
which is independent of the loop integration.
In this remainder the following string of Dirac matrices occurs:
\bq
  \gamma_\nu \gamma_\rho
  \gamma_\mu \gamma_\sigma
  \gamma^\nu.
\eq
If we anti-commute the first Dirac matrix, we can achieve that the two Dirac matrices with index $\nu$ are next
to each other:
\bq
 \gamma_\nu \gamma^\nu.
\eq
In four dimensions this equals $4$ times the unit matrix. What is the value within dimensional
regularisation ? The answer depends on how we treat the Dirac algebra. Does the Dirac algebra remain
in four dimensions or do we also continue the Dirac algebra to $D$ dimensions ?
There are several schemes on the market which treat this issue differently.
To discuss these schemes it is best to look how they treat the momenta and the polarisation vectors
of observed and unobserved particles.
Unobserved particles are particles circulating inside loops or emitted particles not resolved within
a given detector resolution.
The most commonly used schemes are the conventional dimensional
regularisation scheme (CDR) \cite{Collins}, 
where all momenta and all polarisation vectors are taken to be in $D$ dimensions,
the 't Hooft-Veltman scheme (HV) \cite{'tHooft:1972fi,Breitenlohner:1977hr}, 
where the momenta and the helicities of the unobserved particles are $D$ dimensional,
whereas the momenta and the helicities of the observed particles are 4 dimensional,
and the four-dimensional helicity scheme (FD) \cite{Bern:1992aq,Weinzierl:1999xb,Bern:2002zk}, 
where all polarisation vectors are kept in four dimensions, as well
as the momenta of the observed particles. 
Only the momenta of the unobserved particles are continued to $D$ dimensions.

The conventional scheme is mostly used for an analytical calculation of the interference of a one-loop
amplitude with the Born amplitude by using polarisation sums corresponding to $D$ dimensions.
For the calculation of one-loop helicity amplitudes the 't Hooft-Veltman scheme
and the four-dimensional helicity scheme are possible choices.
All schemes have in common, that the propagators appearing in the denominator of the
loop-integrals are continued to $D$ dimensions. They differ how they treat the algebraic
part in the numerator.
In the 't Hooft-Veltman scheme the algebraic part is treated in $D$ dimensions, whereas
in the FD scheme the algebraic part is treated in four dimensions.
It is possible to relate results obtained in one scheme to another scheme, using simple
and universal transition formulae \cite{Kunszt:1994sd,Signer:PhD,Catani:1997pk}.
Therefore, if we return to the example above, we have
\bq
 \gamma_\nu \gamma^\nu
 & = & 
 \left\{ \begin{array}{ll}
         D \cdot {\bf 1}, & \mbox{in the CDR and HV scheme,} \\
         4 \cdot {\bf 1}, & \mbox{in the FD scheme.} \\
         \end{array}
 \right.
\eq
To summarise we are interested into loop integrals regulated by dimensional regularisation.
As a result we seek the Laurent expansion around $D=4$. It is common practise to parametrise the
deviation of $D$ from $4$ by
\bq 
 D & = & 4 - 2\eps.
\eq
Divergent loop integrals will therefore have poles in $1/\eps$. 
In an $l$-loop integral ultraviolet divergences will lead to poles $1/\eps^l$ at the worst, whereas
infrared divergences can lead to poles up to $1/\eps^{2l}$.

\section{Loop integration in $D$ dimensions}
\label{sect:loopint}

In this section I will discuss how to perform the $D$-dimensional loop integrals.
It would be more correct to say that we exchange them for some parameter integrals.
As an example we take the one-loop integral of eq.~(\ref{loop_int_example_1a}):
\bq
\label{basic_scalar_int}
 I & = &
 \int \frac{d^Dk_1}{i \pi^{D/2}}
 \frac{1}{(-k_1^2) (-k_2^2) (-k_3^2)}
\eq
The integration is now in $D$ dimensions.
In eq.~(\ref{basic_scalar_int}) there are some overall factors, which I inserted for convenience:
The integral measure is now $d^Dk/(i \pi^{D/2})$ instead of $d^Dk/(2 \pi)^D$, and each propagator
is multiplied by $(-1)$. The reason for doing this is that the final result will be simpler.

As already discussed above, the only functions we really want to integrate over $D$ dimensions
are the ones which depend on the loop momentum only through $k^2$. The integrand in 
eq.~(\ref{basic_scalar_int}) is not yet in such a form.
To bring the integrand into this form, we first convert the product of propagators into 
a sum. 
We can do this with the Feynman parameter technique.
In its full generality it is also applicable to cases, where each factor in the denominator is raised to 
some power $\nu$.
The formula reads:
\bq
 \prod\limits_{i=1}^{n} \frac{1}{\left(-P_{i}\right)^{\nu_{i}}} 
 & = &
 \frac{\Gamma(\nu)}{\prod\limits_{i=1}^{n} \Gamma(\nu_{i})}
 \int\limits_{0}^{1} \left( \prod\limits_{i=1}^{n} dx_{i} \; x_{i}^{\nu_{i}-1} \right)
 \frac{\delta\left(1-\sum\limits_{i=1}^{n} x_{i}\right)}
      {\left( - \sum\limits_{i=1}^{n} x_{i} P_{i} \right)^{\nu}}, 
\;\;\;\;\;\;
 \nu = \sum\limits_{i=1}^{n} \nu_{i}. 
\eq
The proof of this formula can be found in many text books and is not repeated here.
$\Gamma(x)$ is Euler's Gamma function, $\delta(x)$ denotes Dirac's delta function.
The price we have to pay for converting the product into a sum are $(n-1)$ additional integrations.
Let us look at the example from eq.~(\ref{loop_int_example_1a}):
\bq
\label{example_feynman_parameterisation}
 \frac{1}{(-k_1^2) (-k_2^2) (-k_3^2)}
 & = &
 2 \int\limits_{0}^{1} dx_1 \int\limits_{0}^{1} dx_2 \int\limits_{0}^{1} dx_3 
 \frac{\delta(1-x_1-x_2-x_3)}{ \left( -x_1 k_1^2 - x_2 k_2^2 - x_3 k_3^2 \right)^{3}}.
\eq
In the next step we complete the square and shift the loop momentum, such that 
the integrand becomes a function of $k^2$.
With $k_2=k_1-p_{12}$ and  $k_3=k_2-p_3$ we have
\bq
 -x_1 k_1^2 - x_2 k_2^2 - x_3 k_3^2
 & = & - \left( k_1 - x_2 p_{12} - x_3p_{123} \right)^2 
       - x_1 x_2 s_{12} - x_1 x_3 s_{123},
\;\;\;\;\;\;\;\;
\eq
where $s_{12}=(p_1+p_2)^2$ and $s_{123}=(p_1+p_2+p_3)^2$.
We can now define 
\bq
 k_1' & = & k_1 - x_2 p_{12} - x_3p_{123} 
\eq
and using translational invariance our loop integral becomes
\bq
\label{example_shift}
 I & = &
 2 \int \frac{d^Dk_1'}{i \pi^{D/2}}
 \int\limits_{0}^{1} dx_1 \int\limits_{0}^{1} dx_2 \int\limits_{0}^{1} dx_3 
 \frac{\delta(1-x_1-x_2-x_3)}{ \left( - {k_1'}^2 - x_1 x_2 s_{12} - x_1 x_3 s_{123} \right)^{3}}.
\eq
The integrand is now a function of ${k_1'}^2$, which we can relabel as $k^2$.

Having succeeded to rewrite the integrand as a function of $k^2$, we then perform a Wick rotation, which
transforms Minkowski space into an Euclidean space.
Remember, that $k^2$ written out in components in $D$-dimensional Minkowski space reads
\bq
 k^2 = k_0^2 - k_1^2 - k_2^2 - k_3^2 - ...
\eq
(Here $k_j$ denotes the $j$-th component of the vector $k$, in contrast to the previous notation, where
we used the subscript to label different vectors $k_j$. It should be clear from the context what is meant.)
Furthermore, when integrating over $k_0$, we encounter poles which are avoided by Feynman's 
$i\delta$-prescription.
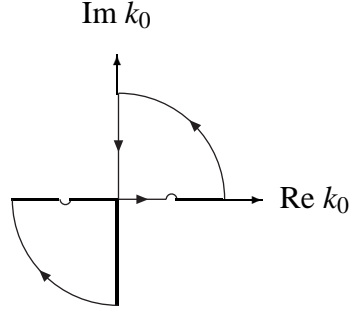
\begin{figure}
\begin{center}
\begin{picture}(300,125)(0,0)
\thicklines
\put(110,50){\line(1,0){18}}
\CArc(130,50)(2,180,0)
\put(132,50){\line(1,0){18}}
\ArrowLine(150,50)(168,50)
\CArc(170,50)(2,0,180)
\put(172,50){\line(1,0){18}}
\ArrowLine(150,90)(150,50)
\put(150,50){\line(0,-1){40}}
\ArrowArc(150,50)(40,0,90)
\ArrowArcn(150,50)(40,270,180)
\thinlines
\put(190,50){\vector(1,0){15}}
\put(210,40){\makebox(30,20){Re $k_{0}$}}
\put(150,90){\vector(0,1){15}}
\put(135,110){\makebox(30,20){Im $k_{0}$}}
\end{picture}
\end{center}
\caption{\label{fig2} Integration contour for the Wick rotation. The little circles along the real axis
exclude the poles.}
\end{figure}    
In the complex $k_0$-plane we consider the integration contour shown in fig.~\ref{fig2}.
Since the contour does not enclose any poles, the integral along the complete contour is zero:
\bq
\oint dk_{0} f(k_0) & = & 0.
\eq
If the quarter-circles at infinity give a vanishing contribution
(it can be shown that this is the case)
we obtain
\bq
\label{wick_rotation}
\int\limits_{-\infty}^{\infty} dk_{0} f(k_0) 
 & = & - \int\limits_{i \infty}^{-i \infty} dk_{0} f(k_0).
\eq
We now make the following change of variables:
\bq
\label{change_variables_wick_rotation}
 k_{0} & = & i K_{0}, \nonumber \\
 k_j   & = & K_j, \;\;\;\;\;\mbox{for}\; 1 \le j \le D-1.
\eq
As a consequence we have
\bq
k^{2} = - K^{2}, 
 & &
d^{D}k = i d^{D}K,
\eq
where $K^2$ is now given with Euclidean signature:
\bq
 K^2 & = & K_0^2 + K_1^2 + K_2^2 + K_3^2 + ...
\eq
Combining eq.~(\ref{wick_rotation}) with eq.~(\ref{change_variables_wick_rotation})
we obtain for the integration of a function $f(k^2)$ over $D$ dimensions
\bq
\label{final_wick_rotation}
 \int \frac{d^Dk}{i \pi^{D/2}} f(-k^2)
 & = & 
 \int \frac{d^DK}{\pi^{D/2}} f(K^2),
\eq
whenever there are no poles inside the contour of fig.~\ref{fig2} and the arcs at infinity give a 
vanishing contribution.
The integral on the r.h.s. is now over $D$-dimensional Euclidean space.
Eq.~(\ref{final_wick_rotation}) justifies our conventions, to introduce a factor $i$ in the denominator
and a minus sign for each propagator in eq.~(\ref{basic_scalar_int}).
These conventions are just such that after Wick rotation we have simple formulae.

We now have an integral over $D$-dimen\-sional Euclidean space, where the integrand depends only on
$K^2$. It is therefore natural to introduce spherical coordinates. In $D$ dimensions they are given by
\bq
 K_{0} & = & K \cos \theta_{1}, \nonumber \\
 K_{1} & = & K \sin \theta_{1} \cos \theta_{2}, \nonumber \\
 & ... & \nonumber \\
 K_{D-2} & = & K \sin \theta_{1} ... \sin \theta_{D-2} \cos \theta_{D-1}, \nonumber \\
 K_{D-1} & = & K \sin \theta_{1} ... \sin \theta_{D-2} \sin \theta_{D-1}.
\eq
In $D$ dimensions we have one radial variable $K$, $D-2$ polar angles $\theta_j$ (with $1 \le j \le D-2$)
and one azimuthal angle $\theta_{D-1}$.
The measure becomes
\bq
d^{D}K & = & K^{D-1} dK d\Omega_{D},
 \;\;\;\;\;\;
 d\Omega_{D} = \prod\limits_{i=1}^{D-1} \sin^{D-1-i} \theta_{i} \; d\theta_{i}.
\eq
Integration over the angles yields
\bq
\label{angular_integration}
 \int d\Omega_{D} & = & \int\limits_{0}^{\pi} d\theta_{1} \sin^{D-2} \theta_{1}
 ... \int\limits_{0}^{\pi} d\theta_{D-2} \sin \theta_{D-2} 
 \int\limits_{0}^{2 \pi} d\theta_{D-1} 
 = \frac{2 \pi^{D/2}}{\Gamma\left( \frac{D}{2} \right)}.
\eq
Note that the integration on the l.h.s
of eq.~(\ref{angular_integration}) is defined for any natural number $D$, whereas the result
on the r.h.s is an analytic function of $D$, which can be continued to any complex value.

It is now the appropriate place to say a few words on Euler's Gamma function.
The Gamma function is defined for $\mbox{Re}(x) > 0$ by
\bq
\Gamma(x) & = & \int_{0}^{\infty} e^{-t} t^{x-1} dt.
\eq
It fulfils the functional equation
\bq
\Gamma(x+1) & = & x \; \Gamma(x).
\eq
For positive integers $n$ it takes the values
\bq
\Gamma(n+1) & = & n! = 1 \cdot 2 \cdot 3 \cdot ... \cdot n.
\eq
For integers $n$ we have the reflection identity
\bq
\frac{\Gamma(x-n)}{\Gamma(x)} & = & \left(-1 \right)^n \frac{\Gamma(1-x)}{\Gamma(1-x+n)}.
\eq
The Gamma function $\Gamma(x)$ has poles located on the negative real axis at $x=0,-1,-2,...$.
Quite often we will need the expansion around these poles.
This can be obtained from the expansion around $x=1$ and the functional equation.
The expansion around $\eps=1$ reads
\bq
\Gamma(1+\eps)  & = & 
  \exp \left( - \gamma_E \eps + \sum\limits_{n=2}^\infty \frac{(-1)^n}{n} \zeta_n \eps^n \right),
\eq
where
$\gamma_E$ is Euler's constant
\bq 
 \gamma_E & = & \lim\limits_{n\rightarrow \infty} \left( \sum\limits_{j=1}^n \frac{1}{j} - \ln n \right)
 = 0.5772156649...
\eq
and $\zeta_n$ is given by
\bq
 \zeta_n & = & \sum\limits_{j=1}^\infty \frac{1}{j^n}.
\eq
For example we obtain for the Laurent expansion around $\eps=0$
\bq
\Gamma(\varepsilon) = \frac{1}{\varepsilon} - \gamma_E + O(\varepsilon).
\eq 
We are now in a position to perform the integration over the loop momentum.
Let us discuss again the example from eq.~(\ref{example_shift}).
After Wick rotation we have
\bq
 I & = &
 \int \frac{d^Dk_1}{i \pi^{D/2}}
 \frac{1}{(-k_1^2) (-k_2^2) (-k_3^2)}
 = 
 2 \int \frac{d^DK}{\pi^{D/2}}
 \int d^3x 
 \frac{\delta(1-x_1-x_2-x_3)}{ \left( K^2 - x_1 x_2 s_{12} - x_1 x_3 s_{123} \right)^{3}}.
 \nonumber \\
\eq
Introducing spherical coordinates and performing the angular integration this becomes
\bq
 I & = &
 \frac{2}{\Gamma\left(\frac{D}{2}\right)} \int\limits_0^\infty dK^2
 \int d^3x 
 \frac{\delta(1-x_1-x_2-x_3) \left(K^2\right)^{\frac{D-2}{2}}}{ \left( K^2 - x_1 x_2 s_{12} - x_1 x_3 s_{123} \right)^{3}}.
\eq
For the radial integration we have after the substitution $t=K^2/(- x_1 x_2 s_{12} - x_1 x_3 s_{123})$
\bq
 \int\limits_0^\infty dK^2
 \frac{\left(K^2\right)^{\frac{D-2}{2}}}{ \left( K^2 - x_1 x_2 s_{12} - x_1 x_3 s_{123} \right)^{3}}
 & = &
 \left( - x_1 x_2 s_{12} - x_1 x_3 s_{123} \right)^{\frac{D}{2}-3}
 \int\limits_0^\infty dt
 \frac{t^{\frac{D-2}{2}}}{ \left( 1+t \right)^{3}}.
 \nonumber \\
\eq
The remaining integral is a standard integral and yields
\bq
 \int\limits_0^\infty dt
 \frac{t^{\frac{D-2}{2}}}{ \left( 1+t \right)^{3}}
 & = & \frac{\Gamma\left(\frac{D}{2}\right) \Gamma\left(3-\frac{D}{2}\right)}{\Gamma(3)}.
\eq
Putting everything together and setting $D=4-2\eps$ we obtain
\bq
\label{example_in_feynman_parameters}
 I
 = 
 \Gamma\left(1+\eps\right)
 \int d^3x \; \delta(1-x_1-x_2-x_3) \; x_1^{-1-\eps}
 \left( - x_2 s_{12} - x_3 s_{123} \right)^{-1-\eps}.
\eq
Therefore we succeeded in performing the integration over the loop momentum $k$ at the expense of 
introducing a two-fold integral over the Feynman parameters.
We will learn techniques how to perform the Feynman parameter integrals later in these lectures.
Let me however already state the final result:
\bq
\label{final_result}
 I & = & - \frac{1}{s_{123}-s_{12}} \left[ \left( \frac{1}{\eps} - \gamma_E - \ln\left(-s_{123}\right) \right) \ln x - \frac{1}{2} \ln^2 x \right]
 + {\cal O}(\eps),
 \;\;\;
 x = \frac{-s_{12}}{-s_{123}}.
\eq
The result has been expanded in the regularisation parameter $\eps$ up to the order ${\cal O}(\eps)$. We see that the result
has a term proportional to $1/\eps$. Poles in $\eps$ in the final (regularised) result reflect the original divergences in the
unregularised integral. In this example the pole corresponds to a collinear singularity.

\section{Multi-loop integrals}
\label{sect:multi_loop}

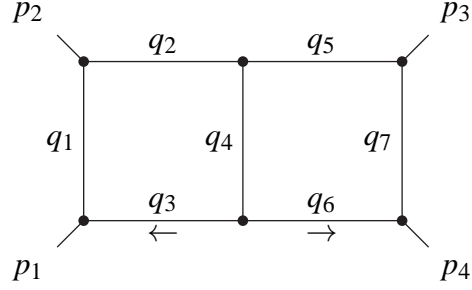
\begin{figure}
\begin{center}
\begin{picture}(200,100)(0,0)
 \Line(10,10)(20,20)
 \Line(10,90)(20,80)
 \Vertex(20,20){2}
 \Vertex(20,80){2}
 \Line(20,20)(20,80)
 \Line(20,20)(140,20)
 \Line(20,80)(140,80)
 \Line(80,20)(80,80)
 \Line(140,20)(140,80)
 \Vertex(80,20){2}
 \Vertex(80,80){2}
 \Vertex(140,20){2}
 \Vertex(140,80){2}
 \Line(140,20)(150,10)
 \Line(140,80)(150,90)
 \Text(5,5)[rt]{$p_1$}
 \Text(5,95)[rb]{$p_2$}
 \Text(155,5)[lt]{$p_4$}
 \Text(155,95)[lb]{$p_3$}
 \Text(50,23)[b]{$q_3$}
 \Text(110,23)[b]{$q_6$}
 \Text(50,83)[b]{$q_2$}
 \Text(110,83)[b]{$q_5$}
 \Text(17,50)[r]{$q_1$}
 \Text(77,50)[r]{$q_4$}
 \Text(137,50)[r]{$q_7$}
 \Text(50,17)[t]{$\leftarrow$}
 \Text(110,17)[t]{$\rightarrow$}
\end{picture}
\end{center}
\caption{\label{figdoublebox} The ``double box''-graph:
A two-loop Feynman diagram with four external lines and seven internal lines.
The momenta flowing out along the external lines are labelled $p_1$, ..., $p_4$, 
the momenta flowing through the internal lines are labelled $q_1$, ..., $q_7$.}
\end{figure}

As the steps discussed in the previous section always occur in any loop integration we can combine them into a master formula.
Let us consider a scalar Feynman graph $G$ with $m$ external lines and $n$ internal lines.
We denote by $I_G$ the associated scalar $l$-loop integral.
For each internal line $j$ the corresponding propagator
in the integrand can be raised to an integer power $\nu_j$.
Therefore the integral will depend also on the numbers $\nu_1$,...,$\nu_n$.
\bq
\label{eq_basic_feynman_integral}
I_G  & = &
 \int \prod\limits_{r=1}^{l} \frac{d^Dk_r}{i\pi^{\frac{D}{2}}}\;
 \prod\limits_{j=1}^{n} \frac{1}{(-q_j^2+m_j^2)^{\nu_j}}.
\eq
The independent loop momenta are labelled $k_1$, ..., $k_l$.
The momenta flowing through the propagators are then given
as a linear combination of the external 
momenta $p$ and the loop momenta $k$ with coefficients $-1$, $0$ or $1$:
\bq
\label{eq_internal_mom}
 q_i & = & \sum\limits_{j=1}^l \lambda_{ij} k_j + \sum\limits_{j=1}^m \sigma_{ij} p_j,
 \;\;\; \lambda_{ij}, \sigma_{ij} \in \{-1,0,1\}.
\eq
We can repeat for each loop integration the steps of the previous section. Doing so, we arrive 
at the following Feynman parameter integral:
\bq
\label{eq_feynman_parameter_integral}
I_G  & = &
 \frac{\Gamma(\nu-lD/2)}{\prod\limits_{j=1}^{n}\Gamma(\nu_j)}
 \int\limits_{x_j \ge 0} d^nx \; \delta(1-\sum_{i=1}^n x_i)
 \left( \prod\limits_{j=1}^{n}\,dx_j\,x_j^{\nu_j-1} \right)\,\frac{{\mathcal U}^{\nu-(l+1) D/2}}
 {{\mathcal F}^{\nu-l D/2}}.
\eq
The functions ${\mathcal U}$ and $\mathcal F$ depend on the Feynman parameters $x_j$.
If one expresses
\bq
\label{eq_poly_calc_1}
 \sum\limits_{j=1}^{n} x_{j} (-q_j^2+m_j^2)
 & = & 
 - \sum\limits_{r=1}^{l} \sum\limits_{s=1}^{l} k_r M_{rs} k_s + \sum\limits_{r=1}^{l} 2 k_r \cdot Q_r + J,
\eq
where $M$ is a $l \times l$ matrix with scalar entries and $Q$ is a $l$-vector
with four-vectors as entries,
one obtains
\bq
\label{eq_poly_calc_2}
 {\mathcal U} = \mbox{det}(M),
 & &
 {\mathcal F} = \mbox{det}(M) \left( J + Q M^{-1} Q \right).
\eq
As an example let us look at the two-loop double box graph of fig.~(\ref{figdoublebox}).
In fig.~\ref{figdoublebox} there are two independent loop momenta. 
We can choose them to be $k_1=q_3$ and $k_2=q_6$.
Then all other internal momenta are expressed in terms of $k_1$, $k_2$ and the external momenta
$p_1$, ..., $p_4$:
\bq
\begin{array}{lll}
 q_1 = k_1 - p_1,
&
 q_2 = k_1 - p_1 - p_2,
&
 q_4 = k_1 + k_2,
 \\
 q_5 = k_2 - p_3 - p_4,
 &
 q_7 = k_2 - p_4.
 & \\
\end{array}
\eq
We will consider the case
\bq
\label{specification_double_box}
 & & p_1^2 = 0, \;\;\; p_2^2 = 0, \;\;\; p_3^2 = 0, \;\;\; p_4^2 = 0,
 \nonumber \\
 & & m_1 = m_2 = m_3 = m_4 = m_5 = m_6 = m_7 = 0.
\eq
We define
\bq
 s = \left(p_1+p_2\right)^2=\left(p_3+p_4\right)^2,
 & &
 t = \left(p_2+p_3\right)^2=\left(p_1+p_4\right)^2.
\eq
We have
\bq
 \sum\limits_{j=1}^7 x_j \left(-q_j^2\right) & = &
 - \left(x_1+x_2+x_3+x_4\right) k_1^2 - 2 x_4 k_1 \cdot k_2 - \left( x_4+x_5+x_6+x_7\right) k_2^2
 \\
 & &
 + 2 \left[ x_1 p_1 + x_2 \left( p_1 + p_2 \right) \right] \cdot k_1
 + 2 \left[ x_5 \left( p_3 + p_4 \right) + x_7 p_4 \right] \cdot k_2
 - \left( x_2 + x_5 \right) s.
 \nonumber
\eq
In comparing with eq.~(\ref{eq_poly_calc_1})
we find
\bq
 M & = & \left( \begin{array}{cc}
 x_1+x_2+x_3+x_4 & x_4 \\
 x_4 & x_4+x_5+x_6+x_7 \\
 \end{array} \right),
 \nonumber \\
 Q & = & \left( \begin{array}{c}
          x_1 p_1 + x_2 \left( p_1 + p_2 \right) \\
          x_5 \left( p_3 + p_4 \right) + x_7 p_4 \\
          \end{array} \right),
 \nonumber \\
 J & = & \left( x_2 + x_5 \right) \left(-s\right).
\eq
Plugging this into eq.~(\ref{eq_poly_calc_2})
we obtain the graph polynomials as
\bq
\label{result_U_and_F_double_box}
{\mathcal U} & = & \left( x_1+x_2+x_3 \right) \left( x_5+x_6+x_7 \right) + x_4 \left( x_1+x_2+x_3+x_5+x_6+x_7 \right),
 \nonumber \\
{\mathcal F} & = & \left[ x_2 x_3 \left( x_4+x_5+x_6+x_7 \right)
                        + x_5 x_6 \left( x_1+x_2+x_3+x_4 \right)
                        + x_2 x_4 x_6 + x_3 x_4 x_5 \right] \left( -s \right)
 \nonumber \\
 & &
      + x_1 x_4 x_7 \left( -t \right).
\eq
There are several other ways how the two polynomials ${\mathcal U}$ and ${\mathcal F}$ can be obtained \cite{Bogner:2010kv}.
Let me mention one method, where the two polynomials can be read off directly from the topology of the
graph $G$. 
We consider first connected tree graphs $T$, which are obtained from the graph $G$ by cutting $l$ lines.
The set of all such trees (or 1-trees) is denoted by ${\mathcal T}_1$.  
The Feynman parameters corresponding to the cut lines define a monomial of degree $l$.
${\mathcal U}$ is the sum over all such monomials.
Cutting one more line of a 1-tree leads to two disconnected trees $(T_1,T_2)$, or a 2-tree.
${\mathcal T}_2$ is the set of all such  pairs.
The cut lines define  monomials of degree $l+1$. Each 2-tree of a graph
corresponds to a cut defined by cutting the lines which connected the two now disconnected trees
in the original graph. 
The square of the sum of momenta through the cut lines 
of one of the two disconnected trees $T_1$ or $T_2$
defines a Lorentz invariant
\bq
s_{(T_1,T_2)} & = & \left( \sum\limits_{j\notin (T_1,T_2)} q_j \right)^2.
\eq   
The function ${\mathcal F}_0$ is the sum over all such monomials times 
minus the corresponding invariant. The function ${\mathcal F}$ is then given by ${\mathcal F}_0$ plus an additional piece
involving the internal masses $m_j$.
In summary, the functions ${\mathcal U}$ and ${\mathcal F}$ are obtained from the graph as follows:
\bq
\label{eq0def}	
 {\mathcal U} 
 & = & 
 \sum\limits_{T\in {\mathcal T}_1} \Bigl[\prod\limits_{j\notin T}x_j\Bigr]\;,
 \\
 {\mathcal F}_0 
 & = & 
 \sum\limits_{(T_1,T_2)\in {\mathcal T}_2}\;\Bigl[ \prod\limits_{j\notin (T_1,T_2)} x_j \Bigr]\, (-s_{(T_1,T_2)})\;,
 \;\;\;\;\;\;
 {\mathcal F} 
 =   
 {\mathcal F}_0 + {\mathcal U} \sum\limits_{j=1}^{n} x_j m_j^2\;.
 \nonumber 
\eq

\section{How to obtain finite results}
\label{sect:finite}

We have already seen in eq.~(\ref{final_result}) that the final result of a regularised Feynman integral may contain
poles in the regularisation parameter $\eps$.
These poles reflect the original ultraviolet and infrared singularities of the unregularised integral.
What shall we do with these poles ? The answer has to come from physics and we distinguish again the case of
UV-divergences and IR-divergences.
The UV-divergences are removed through renormalisation.
Ultraviolet divergences are absorbed into a redefinition of the parameters.
As an example we consider the renormalisation of the coupling:
\bq
 \underbrace{g}_{\mathrm{divergent}} 
 & = & 
 \underbrace{Z_g}_{\mathrm{divergent}} \underbrace{g_r}_{\mathrm{finite}}.
\eq
The renormalisation constant $Z_g$ absorbs the divergent part. However $Z_g$ is not unique: One may always shift
a finite piece from $g_r$ to $Z_g$ or vice versa.
Different choices for $Z_g$ correspond to different renormalisation schemes.
Two different renormalisation schemes are always connected by a finite renormalisation.
Note that different renormalisation schemes give numerically different answers.
Therefore one always has to specify the renormalisation scheme.
Some popular renormalisation schemes are
the on-shell scheme, where the renormalisation constants are defined by conditions at a scale where the particles are on-shell.
A second widely used scheme is modified minimal subtraction.
In this scheme one always absorbs the combination
\bq
 \Delta & = & \frac{1}{\eps} - \gamma_E + \ln 4 \pi
\eq
into the renormalisation constants.
One proceeds similar with all other quantities appearing in the original Lagrangian. For example:
\bq
 A_\mu^a = \sqrt{Z_3} A^a_{\mu,r}, 
 \;\;\;
 \psi_q = \sqrt{Z_2} \psi_{q,r},
 \;\;\;
g = Z_g g_r,
 \;\;\;
m = Z_m m_r,
 \;\;\;
\xi = Z_\xi \xi_r.
\eq
The fact that square roots appear for the field renormalisation is just convention.
Let us look a little bit closer into the coupling renormalisation within dimensional regularisation 
and the $\overline{\mathrm{MS}}$-renormalisation scheme.
Within dimensional regularisation the renormalised coupling $g_r$ is a dimensionfull quantity.
We define a dimensionless quantity $g_R$ by
\bq
g_r & = & g_R \mu^\varepsilon,
\eq
where $\mu$ is an arbitrary mass scale.
From a one-loop calculation one obtains
\bq
\label{Z_g}
Z_g & = & 
 1 -\frac{1}{2} \beta_0 \frac{g_R^2}{(4 \pi)^2} \Delta + {\cal O}(g_R^4),
 \;\;\;
\beta_0 = \frac{11}{3} N_c - \frac{2}{3} N_f.
\eq
$N_c$ is the number of colours and $N_f$ the number of light quarks.
The quantity $g_R$ will depend on the arbitrary scale $\mu$. To derive this dependence one first notes that
the unrenormalised coupling constant $g$ is of course independent of $\mu$:
\bq
\frac{d}{d\mu} g & = & 0
\eq
Substituting $g=Z_g \mu^\eps g_R$ into this equation one obtains
\bq
 \mu \frac{d}{d\mu} g_R
 & = &
 - \eps g_R
 - \left( Z_g^{-1} \mu \frac{d}{d\mu} Z_g \right) g_R.
\eq
From eq.~(\ref{Z_g}) one obtains
\bq
 Z_g^{-1} \mu \frac{d}{d\mu} Z_g 
 & = & 
 \beta_0 \frac{g_R^2}{(4 \pi)^2}
 + {\cal O}(g_R^4).
\eq
Instead of $g_R$ one often uses the quantity $\alpha_s = g_R^2/(4 \pi)$,
Going to $D=4$ one arrives at 
\bq
\mu^2 \frac{d}{d \mu^2} \frac{\alpha_s}{4\pi} 
 & = & 
- \beta_0 \left( \frac{\alpha_s}{4\pi} \right)^2
 + {\cal O}\left( \frac{\alpha_s}{4\pi} \right)^3.
\eq
This differential equation gives the dependence of $\alpha_s$ on the renormalisation scale $\mu$.
At leading order the solution is given by
\bq
\frac{\alpha_s(\mu)}{4 \pi} & = & \frac{1}{\beta_0 \ln \left( \frac{\mu^2}{\Lambda^2} \right)},
\eq
where $\Lambda$ is an integration constant. The quantity $\Lambda$ is called the QCD scale parameter.
For QCD $\beta_0$ is positive and $\alpha_s(\mu)$ decreases with larger $\mu$. This property is called
asymptotic freedom: The coupling becomes smaller at high energies.
In QED $\beta_0$ has the opposite sign and the fine-structure constant $\alpha(\mu)$ increases with larger $\mu$.
The electromagnetic coupling becomes weaker when we go to smaller energies.

Let us now look at the infrared divergences:
We first note that any detector has a finite resolution.
Therefore two particles which are sufficiently close to each other in phase space will be detected as one
particle.
Now let us look again at eqs.~(\ref{observable_master}) and (\ref{basic_perturbative_expansion}).
The next-to-leading order term will receive contributions from the interference term of the one-loop amplitude
${\mathcal A}^{(1)}_n$ with the leading-order amplitude ${\mathcal A}^{(0)}_n$, both with $(n-2)$ final state particles. 
This contribution is of order $g^{2n-2}$. Of the same order is the square of the leading-order amplitude
${\mathcal A}^{(0)}_{n+1}$ with $(n-1)$ final state particles. This contribution we have to take into account whenever
our detector resolves only $n$ particles.
It turns out that the phase space integration over the regions where one or more particles become unresolved
is also divergent, and, when performed in $D$ dimensions, leads to poles with the opposite sign as the one
encountered in the loop amplitudes. Therefore the sum of the two contributions is finite.
The Kinoshita-Lee-Nauenberg theorem 
\cite{Kinoshita:1962ur,Lee:1964is}
guarantees that all infrared divergences cancel, when summed over all
degenerate physical states.
As an example we consider the NLO corrections to $\gamma^\ast \rightarrow 2 \; \mbox{jets}.$
The interference term of the one-loop amplitude with the Born amplitude is given by
\bq
2 \; \mbox{Re} \; \left.{\mathcal A}^{(0)}_3\right.^{\ast} {\mathcal A}^{(1)}_3
 & = & 
 \frac{\alpha_s}{\pi} C_F \left( - \frac{1}{\eps^2} - \frac{3}{2\eps} - 4 + \frac{7}{12} \pi^2 \right)
 S_\eps \left| {\cal A}^{(0)}_3 \right|^2
 + {\cal O}\left(\eps\right).
\eq
$S_\eps=(4\pi)^\eps e^{-\eps\gamma_E}$ is the typical phase-space volume factor in $D=4-2\eps$ dimensions.
For simplicity we have set the renormalisation scale $\mu$ equal to the 
centre-of-mass energy squared $s$.
The square of the Born amplitude is given by
\bq
 \left| {\cal A}^{(0)}_3 \right|^2
 & = & 16 \pi N_c \alpha \left(1-\eps\right) s.
\eq
This is independent of the final state momenta and the integration over the
phase space can be written as
\bq
 \int d\phi_2 \;
\left( 2 \; \mbox{Re} \; \left.{\mathcal A}^{(0)}_3\right.^{\ast} {\mathcal A}^{(1)}_3 \right)
 = 
 \frac{\alpha_s}{\pi} C_F \left( - \frac{1}{\eps^2} - \frac{3}{2\eps} - 4 + \frac{7}{12} \pi^2 \right)
 S_\eps \int d\phi_2 \; \left| {\cal A}^{(0)}_3 \right|^2
 + {\cal O}\left(\eps\right).
 \;\;\;\;\;\;
\eq
The real corrections are given by the leading order matrix element for 
$\gamma^\ast \rightarrow q g \bar{q}$ and read
\bq
 \left| {\cal A}^{(0)}_4 \right|^2 & = & 128 \pi^2 \alpha \alpha_s C_F N_c ( 1 - \varepsilon)  \left[
         \frac{2}{x_1 x_2} 
         - \frac{2}{x_1}
         - \frac{2}{x_2} 
         + (1-\varepsilon) \frac{x_2}{x_1}
         + (1-\varepsilon) \frac{x_1}{x_2} 
         - 2 \varepsilon 
        \right],
 \;\;\;\;\;\;
\eq
where $x_1=s_{12}/s_{123}$, $x_2=s_{23}/s_{123}$ and $s_{123}=s$ is again the centre-of-mass energy squared.
For this particular simple example we can write the three-particle phase space in $D$ dimensions as
\bq
 d\phi_3 & = & d\phi_2 d\phi_{\mathrm{unres}},
 \nonumber \\
d\phi_{\mathrm{unres}} & = & \frac{\left(4\pi\right)^{\eps-2}}{\Gamma\left(1-\eps\right)}
 s_{123}^{1-\eps} d^3x \delta(1-x_1-x_2-x_3) \left( x_1 x_2 x_3 \right)^{-\eps}.
\eq
Integration over the phase space yields
\bq
 \int d\phi_3 \;
 \left| {\cal A}^{(0)}_4 \right|^2 
 & = &
 \frac{\alpha_s}{\pi} C_F \left( \frac{1}{\eps^2} + \frac{3}{2\eps} + \frac{19}{4} - \frac{7}{12} \pi^2 \right)
 S_\eps \int d\phi_2 \; \left| {\cal A}^{(0)}_3 \right|^2
 + {\cal O}\left(\eps\right).
\eq
We see that in the sum the poles cancel and we obtain the finite result
\bq
 \int d\phi_2 \;
\left( 2 \; \mbox{Re} \; \left.{\mathcal A}^{(0)}_3\right.^{\ast} {\mathcal A}^{(1)}_3 \right)
 +
 \int d\phi_3 \;
 \left| {\cal A}^{(0)}_4 \right|^2 
 & = &
 \frac{3}{4} C_F \frac{\alpha_s}{\pi} 
 \int d\phi_2 \; \left| {\cal A}^{(0)}_3 \right|^2
 + {\cal O}\left(\eps\right).
\eq
In this example we have seen the cancellation of the infrared (soft and collinear) singularities 
between the virtual and the real corrections according to the Kinoshita-Lee-Nauenberg theorem.
In this example we integrated over the phase space of all final state particles.
In practise one is often interested in differential distributions.
In these cases the cancellation is technically more complicated, as the different contributions live
on phase spaces of different dimensions and one integrates only over restricted regions of phase space.
Methods to overcome this obstacle are known under the name ``phase-space slicing'' and ``subtraction method''
\cite{Giele:1992vf,Giele:1993dj,Keller:1998tf,Frixione:1996ms,Catani:1997vz,Dittmaier:1999mb,Phaf:2001gc,Catani:2002hc}.

The Kinoshita-Lee-Nauenberg theorem is related to the finite experimental resolution in detecting
final state particles.
In addition we have to discuss initial state particles.
Let us go back to eq.~(\ref{observable_master_hadron}). The differential cross section
we can write schematically
\bq
d \sigma_{H_1 H_2} & = & \sum\limits_{a,b} \int dx_1 f_{H_1 \rightarrow a}(x_1)
\int dx_2 f_{H_2 \rightarrow b}(x_2) d \sigma_{ab}(x_1,x_2),
\eq
where $f_{H \rightarrow a}(x)$ is the parton distribution function, giving us
the probability to find a parton of type $a$ in a hadron of type $H$ carrying a fraction
$x$ to $x+dx$ of the hadron's momentum.
$d \sigma_{ab}(x_1,x_2)$ is the differential cross section for the scattering of partons $a$ and $b$.
Now let us look at the parton distribution function $f_{a \rightarrow b}$ of a parton inside another parton.
At leading order this function is trivially given by $\delta_{ab} \delta(1-x)$, but already at the next order
a parton can radiate off another parton and thus loose some of its momentum and/or convert to another flavour.
One finds in $D$ dimensions
\bq
f_{a \rightarrow b}(x,\eps) & = & \delta_{ab} \delta(1-x)
- \frac{1}{\eps} \frac{\alpha_s}{4 \pi} P^{0}_{a \rightarrow b}(x) 
+ O(\alpha_s^2),
\eq
where $P^{0}_{a \rightarrow b}$ is the lowest order Altarelli-Parisi
splitting function.
To calculate a cross section $d \sigma_{H_1 H_2}$ at NLO involving parton densities one first 
calculates the cross section $d \hat{\sigma}_{a b}$ where the hadrons $H_1$ and $H_2$ are replaced
by partons $a$ and $b$ to NLO:
\bq
d \hat{\sigma}_{a b} & = & d \hat{\sigma}^0_{a b} + \frac{\alpha_s}{4 \pi} d \hat{\sigma}^1_{a b}
+ O(\alpha_s^2)
\eq
The hard scattering part $d \sigma_{a b}$ is then obtained by inserting the perturbative
expansions for $d \hat{\sigma}_{a b}$ and $f_{a \rightarrow b}$ into 
the factorisation formula.
\bq
d \hat{\sigma}^0_{a b} + \frac{\alpha_s}{4 \pi} d \hat{\sigma}^1_{a b} & = &
d \sigma^0_{a b} + \frac{\alpha_s}{4 \pi} d \sigma^1_{a b} 
- \frac{1}{\eps} \frac{\alpha_s}{4 \pi} \sum\limits_c \int dx_1 P^0_{a \rightarrow c} d \sigma^0_{c b}
- \frac{1}{\eps} \frac{\alpha_s}{4 \pi} \sum\limits_d \int dx_2 P^0_{b \rightarrow d} d \sigma^0_{a d}.
 \nonumber
\eq
One therefore obtains for the LO- and the NLO-terms of the hard scattering part
\bq
d \sigma^0_{a b} & = & d \hat{\sigma}^0_{a b} \nonumber \\
d \sigma^1_{a b} & = & d \hat{\sigma}^1_{a b}
+ \frac{1}{\eps} \sum\limits_c \int dx_1 P^0_{a \rightarrow c} d \hat{\sigma}^0_{c b}
+ \frac{1}{\eps} \sum\limits_d \int dx_2 P^0_{b \rightarrow d} d \hat{\sigma}^0_{a d}.
\eq
The last two terms remove the collinear initial state singularities in $d \hat{\sigma}^1_{a b}$.

\section{Feynman integrals and periods}
\label{sect:periods}

In the previous section we have seen how all divergences disappear in the final result.
However in intermediate steps of a calculation we will in general have to deal with expressions
which contain poles in the regularisation parameter $\eps$.
Let us go back to our general Feynman integral as in eq.~(\ref{eq_feynman_parameter_integral}).
We multiply this integral with $e^{l \gamma_E \eps}$, which avoids the occurrence of Euler's constant in the
final result:
\bq
\label{eq_feynman_parameter_integral_repeated}
\hat{I}_G  & = &
 e^{l \gamma_E \eps}
 \frac{\Gamma(\nu-lD/2)}{\prod\limits_{j=1}^{n}\Gamma(\nu_j)}
 \int\limits_{x_j \ge 0} d^nx \; \delta(1-\sum_{i=1}^n x_i)
 \left( \prod\limits_{j=1}^{n}\,dx_j\,x_j^{\nu_j-1} \right)\,\frac{{\mathcal U}^{\nu-(l+1) D/2}}
 {{\mathcal F}^{\nu-l D/2}}.
\eq
This integral has a Laurent series in $\eps$. For a graph with $l$ loops the highest pole of the corresponding 
Laurent series is of power $(2l)$:
\bq
\label{Laurent_expansion}
 \hat{I}_G & = & \sum\limits_{j=-2l}^\infty c_j \eps^j.
\eq
We see that there are three possibilities 
how poles in $\eps$ can arise from the integral in eq.~(\ref{eq_feynman_parameter_integral_repeated}):

First of all the Gamma-function $\Gamma(\nu-l D/2)$ of the prefactor can give rise to 
a (single) pole if the argument of this function is close to zero or to a negative integer
value. This divergence is called the overall ultraviolet divergence.

Secondly, we consider the polynomial ${\mathcal U}$. 
Depending on the exponent $\nu-(l+1)D/2$ of ${\mathcal U}$ 
the vanishing of the polynomial ${\mathcal U}$ 
in some part of the integration region can lead to poles in $\eps$ after integration.
From the definition of ${\mathcal U}$ in eq.~(\ref{eq0def}) one sees that
each term of the expanded form of the polynomial ${\mathcal U}$ 
has coefficient $+1$, therefore ${\mathcal U}$ can only vanish if some of the Feynman
parameters are equal to zero.
In other words, ${\mathcal U}$ is non-zero (and positive) inside the integration region, but
may vanish on the boundary of the integration region.
Poles in $\eps$ resulting from the vanishing of ${\mathcal U}$ are related to
ultraviolet sub-divergences.

Thirdly, we consider the polynomial ${\mathcal F}$. 
In an analytic calculation one often considers the Feynman integral in the Euclidean region.
The Euclidean region is defined as the region, where all invariants
$(p_{i_1}+p_{i_2}+...+p_{i_k})^2$ are negative or zero, and all internal masses are positive or 
zero.
The result in the physical region is then obtained by analytic continuation.
It can be shown that in the Euclidean region the polynomial ${\mathcal F}$ 
is also non-zero (and positive) inside the integration region.
Therefore under the assumption that the external kinematics is within the Euclidean region
the polynomial ${\mathcal F}$ can only vanish on the boundary of the integration region, 
similar to what has been observed for the the polynomial ${\mathcal U}$.
Depending on the exponent $\nu-l D/2$ of ${\mathcal F}$
the vanishing of the polynomial ${\mathcal F}$ on the boundary of
the integration region may lead to poles in $\eps$ after integration.
These poles are related to infrared divergences.

Now let us consider the integral in the Euclidean region and let us further assume 
that all values of kinematical invariants
and masses are given by rational numbers.
Then it can shown that all coefficients $c_j$ in eq.~(\ref{Laurent_expansion}) are periods \cite{Bogner:2007mn}.
I should first say what a period actually is:
There are several equivalent definitions for a period, but probably the most accessible definition is 
the following \cite{Kontsevich:2001}:
A period is a complex number whose real and imaginary parts are values
of absolutely convergent integrals of rational functions with rational coefficients,
over domains in $\mathbb{R}^n$ given by polynomial inequalities with rational coefficients.
The number of periods is a countable set.
Any rational and algebraic number is a period, but there are also transcendental numbers, which are periods.
An example is the number $\pi$, which can be expressed through the integral
\bq
 \pi & = & \iint\limits_{x^2+y^2\le1} dx \; dy.
\eq
The integral on the r.h.s. clearly shows that $\pi$ is a period.
On the other hand, it is conjectured that the basis of the natural logarithm $e$
and Euler's constant $\gamma_E$
are not periods.
Although there are uncountably many numbers, which are not periods, only very recently an example
for a number which is not a period has been found \cite{Yoshinaga:2008}.

The proof that all coefficients in eq.~(\ref{Laurent_expansion}) are periods is constructive \cite{Bogner:2007mn} 
and based on sector decomposition \cite{Hepp:1966eg,Roth:1996pd,Binoth:2000ps,Binoth:2003ak,Bogner:2007cr,Smirnov:2008py}.
The method can be used to compute numerically each coefficient of the Laurent expansion. This is a very
reliable method, but unfortunately also a little bit slow.

\section{Shuffle algebras}
\label{sect:shuffle}

Before we continue the discussion of loop integrals, it is useful to discuss first
shuffle algebras and generalisations thereof from an algebraic viewpoint.
Consider a set of letters $A$. The set $A$ is called the alphabet.
A word is an ordered sequence of letters:
\bq
 w & = & l_1 l_2 ... l_k.
\eq
The word of length zero is denoted by $e$.
Let $K$ be a field and consider the vector space of words over $K$.
A shuffle algebra ${\cal A}$ on the vector space of words is defined by
\bq
\left( l_1 l_2 ... l_k \right) \cdot 
 \left( l_{k+1} ... l_r \right) & = &
 \sum\limits_{\mbox{\tiny shuffles} \; \sigma} l_{\sigma(1)} l_{\sigma(2)} ... l_{\sigma(r)},
\eq
where the sum runs over all permutations $\sigma$, which preserve the relative order
of $1,2,...,k$ and of $k+1,...,r$.
The name ``shuffle algebra'' is related to the analogy of shuffling cards: If a deck of cards
is split into two parts and then shuffled, the relative order within the two individual parts
is conserved.
A shuffle algebra is also known under the name ``mould symmetral'' \cite{Ecalle}.
The empty word $e$ is the unit in this algebra:
\bq
 e \cdot w = w \cdot e = w.
\eq
A recursive definition of the shuffle product is given by
\bq
\label{def_recursive_shuffle}
\left( l_1 l_2 ... l_k \right) \cdot \left( l_{k+1} ... l_r \right) & = &
 l_1 \left[ \left( l_2 ... l_k \right) \cdot \left( l_{k+1} ... l_r \right) \right]
+
 l_{k+1} \left[ \left( l_1 l_2 ... l_k \right) \cdot \left( l_{k+2} ... l_r \right) \right].
\eq
It is well known fact that the shuffle algebra is actually a (non-cocommutative) Hopf algebra \cite{Reutenauer}.
In this context let us briefly review the definitions of a coalgebra, a bialgebra and a Hopf algebra,
which are closely related:
First note that the unit in an algebra can be viewed as a map from $K$ to $A$ and that the multiplication
can be viewed as a map from the tensor product $A \otimes A$ to $A$ (e.g. one takes two elements
from $A$, multiplies them and gets one element out). 

A coalgebra has instead of multiplication and unit the dual structures:
a comultiplication $\Delta$ and a counit $\bar{e}$.
The counit is a map from $A$ to $K$, whereas comultiplication is a map from $A$ to
$A \otimes A$.
Note that comultiplication and counit go in the reverse direction compared to multiplication
and unit.
We will always assume that the comultiplication is coassociative.
The general form of the coproduct is
\bq
\Delta(a) & = & \sum\limits_i a_i^{(1)} \otimes a_i^{(2)},
\eq
where $a_i^{(1)}$ denotes an element of $A$ appearing in the first slot of $A \otimes A$ and
$a_i^{(2)}$ correspondingly denotes an element of $A$ appearing in the second slot.
Sweedler's notation \cite{Sweedler} consists in dropping the dummy index $i$ and the summation symbol:
\bq
\Delta(a) & = & 
a^{(1)} \otimes a^{(2)}
\eq 
The sum is implicitly understood. This is similar to Einstein's summation convention, except
that the dummy summation index $i$ is also dropped. The superscripts ${}^{(1)}$ and ${}^{(2)}$ 
indicate that a sum is involved.

A bialgebra is an algebra and a coalgebra at the same time,
such that the two structures are compatible with each other.
Using Sweedler's notation,
the compatibility between the multiplication and comultiplication is express\-ed as
\bq
\label{bialg}
 \Delta\left( a \cdot b \right)
 & = &
\left( a^{(1)} \cdot b^{(1)} \right)
 \otimes \left( a^{(2)} \cdot b^{(2)} \right).
\eq

A Hopf algebra is a bialgebra with an additional map from $A$ to $A$, called the 
antipode ${\cal S}$, which fulfils
\bq
a^{(1)} \cdot {\cal S}\left( a^{(2)} \right)
=
{\cal S}\left(a^{(1)}\right) \cdot a^{(2)} 
= e \cdot \bar{e}(a).
\eq

With this background at hand we can now state the coproduct, the counit and the antipode for the
shuffle algebra:
The counit $\bar{e}$ is given by:
\bq
\bar{e}\left( e\right) = 1, \;\;\;
& &
\bar{e}\left( l_1 l_2 ... l_n\right) = 0.
\eq
The coproduct $\Delta$ is given by:
\bq
\Delta\left( l_1 l_2 ... l_k \right) 
& = & \sum\limits_{j=0}^k \left( l_{j+1} ... l_k \right) \otimes \left( l_1 ... l_j \right).
\eq
The antipode ${\cal S}$ is given by:
\bq
{\cal S}\left( l_1 l_2 ... l_k \right) & = & (-1)^k \; l_k l_{k-1} ... l_2 l_1.
\eq
The shuffle algebra is generated by the Lyndon words.
If one introduces a lexicographic ordering on the letters of the alphabet
$A$, a Lyndon word is defined by the property
\bq
w < v
\eq
for any sub-words $u$ and $v$ such that $w= u v$.

\begin{figure}
\begin{center}
\begin{picture}(300,65)(0,0)
\put(10,10){\vector(1,0){50}}
\put(10,10){\vector(0,1){50}}
\Text(60,5)[t]{$t_1$}
\Text(5,60)[r]{$t_2$}
\Line(10,50)(50,50)
\Line(50,50)(50,10)
\Line(10,20)(20,10)
\Line(10,30)(30,10)
\Line(10,40)(40,10)
\Line(10,50)(50,10)
\Line(20,50)(50,20)
\Line(30,50)(50,30)
\Line(40,50)(50,40)
\Text(80,30)[c]{$=$}
\put(110,10){\vector(1,0){50}}
\put(110,10){\vector(0,1){50}}
\Text(160,5)[t]{$t_1$}
\Text(105,60)[r]{$t_2$}
\Line(110,10)(150,50)
\Line(150,50)(150,10)
\Line(120,10)(120,20)
\Line(130,10)(130,30)
\Line(140,10)(140,40)
\Text(180,30)[c]{$+$}
\put(210,10){\vector(1,0){50}}
\put(210,10){\vector(0,1){50}}
\Text(260,5)[t]{$t_1$}
\Text(205,60)[r]{$t_2$}
\Line(210,50)(250,50)
\Line(250,50)(210,10)
\Line(210,20)(220,20)
\Line(210,30)(230,30)
\Line(210,40)(240,40)
\end{picture}
\end{center}
\caption{\label{proof_shuffle} Sketch of the proof for the shuffle product of two iterated integrals.
The integral over the square is replaced by two
integrals over the upper and lower triangle.}
\end{figure}
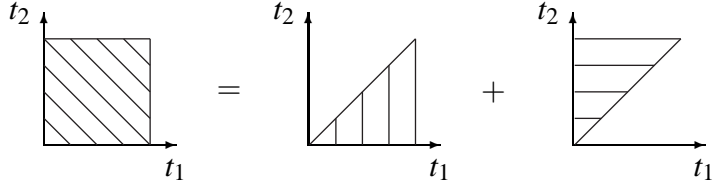

An important example for a shuffle algebra are iterated integrals.
Let $[a, b]$ be a segment of the real line and $f_1$, $f_2$, ... functions
on this interval.
Let us define the following iterated integrals:
\bq
 I(f_1,f_2,...,f_k;a,b) 
 & = &
 \int\limits_a^b dt_1 f_1(t_1) \int\limits_a^{t_1} dt_2 f_2(t_2) 
 ...
 \int\limits_a^{t_{k-1}} dt_k f_k(t_k) 
\eq
For fixed $a$ and $b$ we have a shuffle algebra:
\bq
 I(f_1,f_2,...,f_k;a,b) \cdot I(f_{k+1},...,f_r; a,b) & = &
 \sum\limits_{\mbox{\tiny shuffles} \; \sigma} I(f_{\sigma(1)},f_{\sigma(2)},...,f_{\sigma(r)};a,b),
\eq
where the sum runs over all permutations $\sigma$, which preserve the relative order
of $1,2,...,k$ and of $k+1,...,r$.
The proof is sketched in fig.~\ref{proof_shuffle}.
The two outermost integrations are recursively replaced by integrations over the upper and lower triangle.

We now consider generalisations of shuffle algebras. Assume that for the set of letters we have an additional
operation
\bq
 (.,.) & : & A \otimes A \rightarrow A,
 \nonumber \\
       & &  l_1 \otimes l_2 \rightarrow (l_1, l_2),
\eq
which is commutative and associative.
Then we can define a new product of words recursively through
\bq
\label{def_recursive_quasi_shuffle}
\left( l_1 l_2 ... l_k \right) \ast \left( l_{k+1} ... l_r \right) & = &
 l_1 \left[ \left( l_2 ... l_k \right) \ast \left( l_{k+1} ... l_r \right) \right]
+
 l_{k+1} \left[ \left( l_1 l_2 ... l_k \right) \ast \left( l_{k+2} ... l_r \right) \right]
 \nonumber \\
 & &
+
(l_1,l_{k+1}) \left[ \left( l_2 ... l_k \right) \ast \left( l_{k+2} ... l_r \right) \right].
\eq
This product is a generalisation of the shuffle product and differs from the recursive
definition of the shuffle product in eq.~(\ref{def_recursive_shuffle}) through the extra term in the last line.
This modified product is known under the names quasi-shuffle product \cite{Hoffman},
mixable shuffle product \cite{Guo},
stuffle product \cite{Borwein} or
mould symmetrel \cite{Ecalle}.
Quasi-shuffle algebras are Hopf algebras.
Comultiplication and counit are defined as for the shuffle algebras.
The counit $\bar{e}$ is given by:
\bq
\bar{e}\left( e\right) = 1, \;\;\;
& &
\bar{e}\left( l_1 l_2 ... l_n\right) = 0.
\eq
The coproduct $\Delta$ is given by:
\bq
\Delta\left( l_1 l_2 ... l_k \right) 
& = & \sum\limits_{j=0}^k \left( l_{j+1} ... l_k \right) \otimes \left( l_1 ... l_j \right).
\eq
The antipode ${\cal S}$ is recursively defined through
\bq
{\cal S}\left( l_1 l_2 ... l_k \right) & = & 
 - l_1 l_2 ... l_k
 - \sum\limits_{j=1}^{k-1} {\cal S}\left( l_{j+1} ... l_k \right) \ast \left( l_1 ... l_j \right).
\eq
An example for a quasi-shuffle algebra are nested sums.
Let $n_a$ and $n_b$ be integers with $n_a<n_b$ and let $f_1$, $f_2$, ... be functions
defined on the integers.
We consider the following nested sums:
\bq
 S(f_1,f_2,...,f_k;n_a,n_b) 
 & = &
 \sum\limits_{i_1=n_a}^{n_b} f_1(i_1) \sum\limits_{i_2=n_a}^{i_1-1} f_2(i_2) 
 ...
 \sum\limits_{i_k=n_a}^{i_{k-1}-1} f_k(i_k)
\eq
For fixed $n_a$ and $n_b$ we have a quasi-shuffle algebra:
\bq
\label{quasi_shuffle_multiplication}
\lefteqn{
 S(f_1,f_2,...,f_k;n_a,n_b) \ast S(f_{k+1},...,f_r; n_a,n_b) 
= } & &
 \nonumber \\
 & &
   \sum\limits_{i_1=n_a}^{n_b} f_1(i_1) \; S(f_2,...,f_k;n_a,i_1-1) \ast S(f_{k+1},...,f_r; n_a,i_1-1)
 \nonumber \\
 & &
 +  \sum\limits_{j_1=n_a}^{n_b} f_k(j_1) \; S(f_1,f_2,...,f_k;n_a,j_1-1) \ast S(f_{k+2},...,f_r; n_a,j_1-1)
 \nonumber \\
 & &
 +  \sum\limits_{i=n_a}^{n_b} f_1(i) f_k(i) \; S(f_2,...,f_k;n_a,i-1) \ast S(f_{k+2},...,f_r; n_a,i-1)
\eq
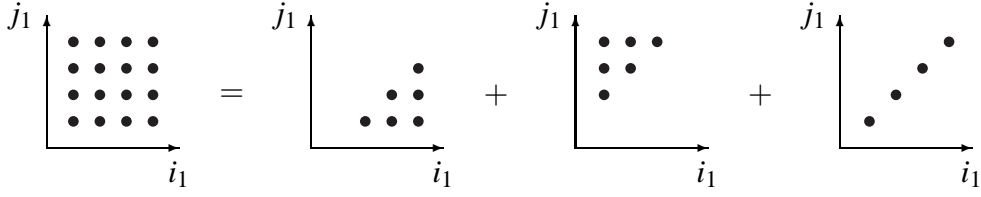
\begin{figure}
\begin{center}
\begin{picture}(400,65)(0,0)
\put(10,10){\vector(1,0){50}}
\put(10,10){\vector(0,1){50}}
\Text(60,5)[t]{$i_1$}
\Text(5,60)[r]{$j_1$}
\Vertex(20,20){2}
\Vertex(30,20){2}
\Vertex(40,20){2}
\Vertex(50,20){2}
\Vertex(20,30){2}
\Vertex(30,30){2}
\Vertex(40,30){2}
\Vertex(50,30){2}
\Vertex(20,40){2}
\Vertex(30,40){2}
\Vertex(40,40){2}
\Vertex(50,40){2}
\Vertex(20,50){2}
\Vertex(30,50){2}
\Vertex(40,50){2}
\Vertex(50,50){2}
\Text(80,30)[c]{$=$}
\put(110,10){\vector(1,0){50}}
\put(110,10){\vector(0,1){50}}
\Text(160,5)[t]{$i_1$}
\Text(105,60)[r]{$j_1$}
\Vertex(130,20){2}
\Vertex(140,20){2}
\Vertex(150,20){2}
\Vertex(140,30){2}
\Vertex(150,30){2}
\Vertex(150,40){2}
\Text(180,30)[c]{$+$}
\put(210,10){\vector(1,0){50}}
\put(210,10){\vector(0,1){50}}
\Text(260,5)[t]{$i_1$}
\Text(205,60)[r]{$j_1$}
\Vertex(220,30){2}
\Vertex(220,40){2}
\Vertex(230,40){2}
\Vertex(220,50){2}
\Vertex(230,50){2}
\Vertex(240,50){2}
\Text(280,30)[c]{$+$}
\put(310,10){\vector(1,0){50}}
\put(310,10){\vector(0,1){50}}
\Text(360,5)[t]{$i_1$}
\Text(305,60)[r]{$j_1$}
\Vertex(320,20){2}
\Vertex(330,30){2}
\Vertex(340,40){2}
\Vertex(350,50){2}
\end{picture}
\end{center}
\caption{\label{proof} Sketch of the proof for the quasi-shuffle product of nested sums. 
The sum over the square is replaced by
the sum over the three regions on the r.h.s.}
\end{figure}
Note that the product of two letters corresponds to the point-wise product of the two functions:
\bq
 ( f_i, f_j ) \; (n) & = & f_i(n) f_j(n).
\eq
The proof that nested sums obey the quasi-shuffle algebra is sketched in Fig. \ref{proof}.
The outermost sums of the nested sums on the l.h.s of (\ref{quasi_shuffle_multiplication}) are split into the three
regions indicated in Fig. \ref{proof}.

\section{Multiple polylogarithms}
\label{sect:polylog}

In the previous section we have seen that iterated integrals form a shuffle algebra, while
nested sums form a quasi-shuffle algebra.
In this context multiple polylogarithms form an interesting class of functions.
They have a representation as iterated integrals as well as nested sums.
Therefore multiple polylogarithms form a shuffle algebra as well as a quasi-shuffle algebra.
The two algebra structures are independent.
Let us start with the representation as nested sums.
The multiple polylogarithms are defined by \cite{Goncharov,Minh:2000,Cartier:2001,Racinet:2002}
\bq 
\label{multipolylog2}
 \mbox{Li}_{m_1,...,m_k}(x_1,...,x_k)
  & = & \sum\limits_{i_1>i_2>\ldots>i_k>0}
     \frac{x_1^{i_1}}{{i_1}^{m_1}}\ldots \frac{x_k^{i_k}}{{i_k}^{m_k}}.
\eq
The multiple polylogarithms are generalisations of
the classical polylogarithms 
$\mbox{Li}_n(x)$,
whose most prominent examples are
\bq
 \mbox{Li}_1(x) = \sum\limits_{i_1=1}^\infty \frac{x^{i_1}}{i_1} = -\ln(1-x),
 & &
 \mbox{Li}_2(x) = \sum\limits_{i_1=1}^\infty \frac{x^{i_1}}{i_1^2},
\eq 
as well as
Nielsen's generalised polylogarithms \cite{Nielsen}
\bq
S_{n,p}(x) & = & \mbox{Li}_{n+1,1,...,1}(x,\underbrace{1,...,1}_{p-1}),
\eq
and the harmonic polylogarithms \cite{Remiddi:1999ew,Gehrmann:2000zt}
\bq
\label{harmpolylog}
H_{m_1,...,m_k}(x) & = & \mbox{Li}_{m_1,...,m_k}(x,\underbrace{1,...,1}_{k-1}).
\eq

In addition, multiple polylogarithms have an integral representation. 
To discuss the integral representation it is convenient to 
introduce for $z_k \neq 0$
the following functions
\bq
\label{Gfuncdef}
G(z_1,...,z_k;y) & = &
 \int\limits_0^y \frac{dt_1}{t_1-z_1}
 \int\limits_0^{t_1} \frac{dt_2}{t_2-z_2} ...
 \int\limits_0^{t_{k-1}} \frac{dt_k}{t_k-z_k}.
\eq
In this definition 
one variable is redundant due to the following scaling relation:
\bq
G(z_1,...,z_k;y) & = & G(x z_1, ..., x z_k; x y)
\eq
If one further defines
\bq
g(z;y) & = & \frac{1}{y-z},
\eq
then one has
\bq
\label{derivative}
\frac{d}{dy} G(z_1,...,z_k;y) & = & g(z_1;y) G(z_2,...,z_k;y)
\eq
and
\bq
\label{Grecursive}
G(z_1,z_2,...,z_k;y) & = & \int\limits_0^y dt \; g(z_1;t) G(z_2,...,z_k;t).
\eq
One can slightly enlarge the set and define
$G(0,...,0;y)$ with $k$ zeros for $z_1$ to $z_k$ to be
\bq
\label{trailingzeros}
G(0,...,0;y) & = & \frac{1}{k!} \left( \ln y \right)^k.
\eq
This permits us to allow trailing zeros in the sequence
$(z_1,...,z_k)$ by defining the function $G$ with trailing zeros via (\ref{Grecursive}) 
and (\ref{trailingzeros}).
To relate the multiple polylogarithms to the functions $G$ it is convenient to introduce
the following short-hand notation:
\bq
\label{Gshorthand}
G_{m_1,...,m_k}(z_1,...,z_k;y)
 & = &
 G(\underbrace{0,...,0}_{m_1-1},z_1,...,z_{k-1},\underbrace{0...,0}_{m_k-1},z_k;y)
\eq
Here, all $z_j$ for $j=1,...,k$ are assumed to be non-zero.
One then finds
\bq
\label{Gintrepdef}
\mbox{Li}_{m_1,...,m_k}(x_1,...,x_k)
& = & (-1)^k 
 G_{m_1,...,m_k}\left( \frac{1}{x_1}, \frac{1}{x_1 x_2}, ..., \frac{1}{x_1...x_k};1 \right).
\eq
The inverse formula reads
\bq
G_{m_1,...,m_k}(z_1,...,z_k;y) & = & 
 (-1)^k \; \mbox{Li}_{m_1,...,m_k}\left(\frac{y}{z_1}, \frac{z_1}{z_2}, ..., \frac{z_{k-1}}{z_k}\right).
\eq
Eq. (\ref{Gintrepdef}) together with 
(\ref{Gshorthand}) and (\ref{Gfuncdef})
defines an integral representation for the multiple polylogarithms.

Up to now we treated multiple polylogarithms from an algebraic point of view.
Equally important are the analytical properties, which are needed for an efficient numerical 
evaluation.
As an example I first discuss the numerical evaluation of the dilogarithm \cite{'tHooft:1979xw}:
\bq
\mbox{Li}_{2}(x) & = & - \int\limits_{0}^{x} dt \frac{\ln(1-t)}{t}
 = \sum\limits_{n=1}^{\infty} \frac{x^{n}}{n^{2}}
\eq
The power series expansion can be evaluated numerically, provided $|x| < 1.$
Using the functional equations 
\bq
\mbox{Li}_2(x) & = & -\mbox{Li}_2\left(\frac{1}{x}\right) -\frac{\pi^2}{6} -\frac{1}{2} \left( \ln(-x) \right)^2,
 \nonumber \\
\mbox{Li}_2(x) & = & -\mbox{Li}_2(1-x) + \frac{\pi^2}{6} -\ln(x) \ln(1-x).
\eq
any argument of the dilogarithm can be mapped into the region
$|x| \le 1$ and
$-1 \leq \mbox{Re}(x) \leq 1/2$.
The numerical computation can be accelerated  by using an expansion in $[-\ln(1-x)]$ and the
Bernoulli numbers $B_i$:
\bq
\mbox{Li}_2(x) & = & \sum\limits_{i=0}^\infty \frac{B_i}{(i+1)!} \left( - \ln(1-x) \right)^{i+1}.
\eq
The generalisation to multiple polylogarithms proceeds along the same lines \cite{Vollinga:2004sn}:
Using the integral representation eq.~(\ref{Gfuncdef})
one transforms all arguments into a region, 
where one has a converging power series expansion.
In this region eq.~(\ref{multipolylog2}) may be used.
However it is advantageous to speed up the convergence of the power series expansion.
This is done as follows: 
The multiple polylogarithms satisfy the H\"older convolution \cite{Borwein}.
For $z_1 \neq 1$ and $z_w \neq 0$ this identity reads
\bq
\label{defhoelder}
\lefteqn{
G\left(z_1,...,z_w; 1 \right) 
 = } & & 
 \\
 & &
 \sum\limits_{j=0}^w \left(-1\right)^j 
  G\left(1-z_j, 1-z_{j-1},...,1-z_1; 1 - \frac{1}{p} \right)
  G\left( z_{j+1},..., z_w; \frac{1}{p} \right).
 \nonumber 
\eq
The H\"older convolution can be used to accelerate the 
convergence for the series
representation of the multiple polylogarithms.

\section{From Feynman integrals to multiple polylogarithms}
\label{sect:calc}

In sect.~\ref{sect:multi_loop} we saw that the Feynman parameter integrals 
depend on two graph polynomials ${\mathcal U}$ and ${\mathcal F}$, which are homogeneous functions of the 
Feynman parameters.
In this section we will discuss how multiple polylogarithms arise in the calculation of Feynman parameter
integrals.
We will discuss two approaches. In the first approach one uses a Mellin-Barnes transformation and sums
up residues. This leads to the sum representation of multiple polylogarithms.
In the second approach one first derives a differential equation for the Feynman parameter integral, which
is then solved by an ansatz in terms of the iterated integral representation of multiple polylogarithms.

Let us start with the first approach. Assume for the moment that the two graph polynomials 
${\mathcal U}$ and ${\mathcal F}$ are absent from the Feynman parameter integral.
In this case we have
\bq
\label{multi_beta_fct}
 \int\limits_{0}^{1} \left( \prod\limits_{j=1}^{n}\,dx_j\,x_j^{\nu_j-1} \right)
 \delta(1-\sum_{i=1}^n x_i)
 & = & 
 \frac{\prod\limits_{j=1}^{n}\Gamma(\nu_j)}{\Gamma(\nu_1+...+\nu_n)}.
\eq
With the help of 
the Mellin-Barnes transformation we now reduce the general case to eq.~(\ref{multi_beta_fct}).
The Mellin-Barnes transformation reads
\bq
\label{multi_mellin_barnes}
\lefteqn{
\left(A_1 + A_2 + ... + A_n \right)^{-c} 
 = 
 \frac{1}{\Gamma(c)} \frac{1}{\left(2\pi i\right)^{n-1}} 
 \int\limits_{-i\infty}^{i\infty} d\sigma_1 ... \int\limits_{-i\infty}^{i\infty} d\sigma_{n-1}
 } & & \\
 & & 
 \times 
 \Gamma(-\sigma_1) ... \Gamma(-\sigma_{n-1}) \Gamma(\sigma_1+...+\sigma_{n-1}+c)
 \; 
 A_1^{\sigma_1} ...  A_{n-1}^{\sigma_{n-1}} A_n^{-\sigma_1-...-\sigma_{n-1}-c}.
 \nonumber 
\eq
Each contour is such that the poles of $\Gamma(-\sigma)$ are to the right and the poles
of $\Gamma(\sigma+c)$ are to the left.
This transformation can be used to convert the sum of monomials of the polynomials ${\mathcal U}$ and ${\mathcal F}$ into
a product, such that all Feynman parameter integrals are of the form of eq.~(\ref{multi_beta_fct}).
As this transformation converts sums into products it is 
the ``inverse'' of Feynman parametrisation.
Eq.~(\ref{multi_mellin_barnes}) is derived from the theory of Mellin transformations:
Let $h(x)$ be a function which is bounded by a power law for $x\rightarrow 0$ and $x \rightarrow \infty$,
e.g.
\bq
\left| h(x) \right| \le K x^{-c_0} & & \mbox{for}\;\; x \rightarrow 0,
 \nonumber \\
\left| h(x) \right| \le K' x^{c_1} & & \mbox{for}\;\; x \rightarrow \infty.
\eq
Then the Mellin transform is defined for
$c_0 < \mbox{Re}\; \sigma < c_1$
by
\bq
h_{\cal M}(\sigma) & = &
 \int\limits_{0}^\infty dx \; h(x) \; x^{\sigma-1}.
\eq
The inverse Mellin transform is given by
\bq
\label{inversemellin}
h(x) & = & \frac{1}{2\pi i} \int\limits_{\gamma-i\infty}^{\gamma+i\infty}
 d\sigma \; h_{\cal M}(\sigma) \; x^{-\sigma}.
\eq
The integration contour is parallel to the imaginary axis and $c_0 < \mbox{Re}\; \gamma < c_1$.
As an example for the Mellin transform we consider the function 
\bq
h(x) & = & \frac{x^c}{(1+x)^c}
\eq
with Mellin transform $h_{\cal M}(\sigma)=\Gamma(-\sigma) \Gamma(\sigma+c) / \Gamma(c)$.
For $\mbox{Re}(-c) < \mbox{Re} \; \gamma < 0$ we have
\bq
\label{baseMellin}
\frac{x^c}{(1+x)^c}
 & = & 
\frac{1}{2\pi i} \int\limits_{\gamma-i\infty}^{\gamma+i\infty}
 d\sigma \; \frac{\Gamma(-\sigma) \Gamma(\sigma+c)}{\Gamma(c)} \; x^{-\sigma}.
\eq
From eq. (\ref{baseMellin}) one obtains with $x=B/A$ the Mellin-Barnes formula
\bq
\label{simple_mellin_barnes}
\left(A+B\right)^{-c}
 & = & 
\frac{1}{2\pi i} \int\limits_{\gamma-i\infty}^{\gamma+i\infty}
 d\sigma \; \frac{\Gamma(-\sigma) \Gamma(\sigma+c)}{\Gamma(c)} \; A^\sigma B^{-\sigma-c}.
\eq
Eq.~(\ref{multi_mellin_barnes}) is then obtained by repeated use of eq.~(\ref{simple_mellin_barnes}).

With the help of eq.~(\ref{multi_beta_fct}) and eq.~(\ref{multi_mellin_barnes})
we may exchange the Feynman parameter integrals against multiple contour integrals.
A single contour integral is of the form
\bq
\label{MellinBarnesInt}
I
 & = & 
\frac{1}{2\pi i} \int\limits_{\gamma-i\infty}^{\gamma+i\infty}
 d\sigma \; 
 \frac{\Gamma(\sigma+a_1) ... \Gamma(\sigma+a_m)}
      {\Gamma(\sigma+c_2) ... \Gamma(\sigma+c_p)}
 \frac{\Gamma(-\sigma+b_1) ... \Gamma(-\sigma+b_n)}
      {\Gamma(-\sigma+d_1) ... \Gamma(-\sigma+d_q)} 
 \; x^{-\sigma}.
\eq
If $\;\mbox{max}\left( \mbox{Re}(-a_1), ..., \mbox{Re}(-a_m) \right) < \mbox{min}\left( \mbox{Re}(b_1), ..., \mbox{Re}(b_n) \right)$ the contour can be chosen
as a straight line parallel to the imaginary axis with
\bq
\mbox{max}\left( \mbox{Re}(-a_1), ..., \mbox{Re}(-a_m) \right) 
 \;\;\; < \;\;\; \mbox{Re} \; \gamma \;\;\; < \;\;\;
\mbox{min}\left( \mbox{Re}(b_1), ..., \mbox{Re}(b_n) \right),
\eq
otherwise the contour is indented, such that the residues of
$\Gamma(\sigma+a_1)$, ..., $\Gamma(\sigma+a_m)$ are to the right of the contour,
whereas the residues of 
$\Gamma(-\sigma+b_1)$,  ..., $\Gamma(-\sigma+b_n)$ are to the left of the contour.
The integral eq. (\ref{MellinBarnesInt}) is most conveniently evaluated with 
the help of the residuum theorem by closing the contour to the left or to the right.
To sum up all residues which lie inside the contour
it is useful to know the residues of the Gamma function:
\bq
\mbox{res} \; \left( \Gamma(\sigma+a), \sigma=-a-n \right) = \frac{(-1)^n}{n!}, 
 & &
\mbox{res} \; \left( \Gamma(-\sigma+a), \sigma=a+n \right) = -\frac{(-1)^n}{n!}. 
\eq
In general there are multiple contour integrals, and as a consequence one obtains multiple sums.
In particular simple cases the contour integrals can be performed in closed form with
the help of two lemmas of Barnes.
Barnes first lemma states that
\bq
\frac{1}{2\pi i} \int\limits_{-i\infty}^{i\infty} d\sigma \;
\Gamma(a+\sigma) \Gamma(b+\sigma) \Gamma(c-\sigma) \Gamma(d-\sigma) 
 =  
\frac{\Gamma(a+c) \Gamma(a+d) \Gamma(b+c) \Gamma(b+d)}{\Gamma(a+b+c+d)},
\eq
if none of the poles of $\Gamma(a+\sigma) \Gamma(b+\sigma)$ coincides with the
ones from $\Gamma(c-\sigma) \Gamma(d-\sigma)$.
Barnes second lemma reads
\bq
\lefteqn{
\frac{1}{2\pi i} \int\limits_{-i\infty}^{i\infty} d\sigma \;
\frac{\Gamma(a+\sigma) \Gamma(b+\sigma) \Gamma(c+\sigma) \Gamma(d-\sigma) \Gamma(e-\sigma)}
{\Gamma(a+b+c+d+e+\sigma)} } & & \nonumber \\
& = & 
\frac{\Gamma(a+d) \Gamma(b+d) \Gamma(c+d) 
      \Gamma(a+e) \Gamma(b+e) \Gamma(c+e)}
{\Gamma(a+b+d+e) \Gamma(a+c+d+e) \Gamma(b+c+d+e)}.
\eq
Although the Mellin-Barnes transformation has been known for a long time, 
the method has seen a revival in applications in recent 
years \cite{Boos:1990rg,Davydychev:1990jt,Davydychev:1990cq,Smirnov:1999gc,Smirnov:1999wz,Tausk:1999vh,Smirnov:2000vy,Smirnov:2000ie,Smirnov:2003vi,Bierenbaum:2003ud,Heinrich:2004iq,Friot:2005cu,Bern:2005iz,Anastasiou:2005cb,Czakon:2005rk,Gluza:2007rt}.

Having collected all residues, one obtains multiple sums.
The task is then to expand all terms in the dimensional regularisation parameter $\eps$ and to re-express
the resulting multiple sums in terms of known functions.
It depends on the form of the multiple sums if this can be done systematically.
The following types of multiple sums occur often and can be evaluated further systematically:
\\
{Type A:}
\bq
\label{type_A}
     \sum\limits_{i=0}^\infty 
       \frac{\Gamma(i+a_1)}{\Gamma(i+a_1')} ...
       \frac{\Gamma(i+a_k)}{\Gamma(i+a_k')}
       \; x^i
\nonumber
\eq
Up to prefactors the hyper-geometric functions ${}_{J+1}F_J$ fall into this class.
\\
{Type B:}
\bq
\label{type_B}
     \sum\limits_{i=0}^\infty 
     \sum\limits_{j=0}^\infty 
       \frac{\Gamma(i+a_1)}{\Gamma(i+a_1')} ...
       \frac{\Gamma(i+a_k)}{\Gamma(i+a_k')}
       \frac{\Gamma(j+b_1)}{\Gamma(j+b_1')} ...
       \frac{\Gamma(j+b_l)}{\Gamma(j+b_l')}
       \frac{\Gamma(i+j+c_1)}{\Gamma(i+j+c_1')} ...
       \frac{\Gamma(i+j+c_m)}{\Gamma(i+j+c_m')}
       \; x^i y^j
\nonumber
\eq
An example for a function of this type is given by the first Appell function $F_1$.
\\
{Type C:}
\bq
\label{type_C}
     \sum\limits_{i=0}^\infty 
     \sum\limits_{j=0}^\infty 
       \left( \begin{array}{c} i+j \\ j \\ \end{array} \right)
       \frac{\Gamma(i+a_1)}{\Gamma(i+a_1')} ...
       \frac{\Gamma(i+a_k)}{\Gamma(i+a_k')}
       \frac{\Gamma(i+j+c_1)}{\Gamma(i+j+c_1')} ...
       \frac{\Gamma(i+j+c_m)}{\Gamma(i+j+c_m')}
       \; x^i y^j
\nonumber
\eq
Here, an example is given by the Kamp\'e de F\'eriet function $S_1$.
\\
{Type D:}
\bq
\label{type_D}
     \sum\limits_{i=0}^\infty 
     \sum\limits_{j=0}^\infty 
       \left( \begin{array}{c} i+j \\ j \\ \end{array} \right)
       \frac{\Gamma(i+a_1)}{\Gamma(i+a_1')} ...
       \frac{\Gamma(i+a_k)}{\Gamma(i+a_k')}
       \frac{\Gamma(j+b_1)}{\Gamma(j+b_1')} ...
       \frac{\Gamma(j+b_l)}{\Gamma(j+b_l')}
       \frac{\Gamma(i+j+c_1)}{\Gamma(i+j+c_1')} ...
       \frac{\Gamma(i+j+c_m)}{\Gamma(i+j+c_m')}
       \; x^i y^j
\nonumber 
\eq
An example for a function of this type is the second Appell function $F_2$.
\\
Note that in these examples there are always as many Gamma functions in the numerator
as in the denominator.
We assume that all $a_n$, $a_n'$, $b_n$, $b_n'$, $c_n$  and $c_n'$ are 
of the form ``integer $+ \;\mbox{const} \cdot \eps$''.
The generalisation towards the form ``rational number $+ \;\mbox{const} \cdot \eps$''
is discussed in \cite{Weinzierl:2004bn}.
The task is now to expand these functions systematically into a Laurent series
in $\eps$.
We start with the formula for the expansion of the Gamma-function:
\bq
\label{expansiongamma}
\lefteqn{
\Gamma(n+\eps)  = 
} & & \\
 & & \Gamma(1+\eps) \Gamma(n)
 \left[
        1 + \eps Z_1(n-1) + \eps^2 Z_{11}(n-1)
          + \eps^3 Z_{111}(n-1) + ... + \eps^{n-1} Z_{11...1}(n-1)
 \right],
 \nonumber
\eq
where $Z_{m_1,...,m_k}(n)$ are Euler-Zagier sums
defined by
\bq
 Z_{m_1,...,m_k}(n) & = &
  \sum\limits_{n \ge i_1>i_2>\ldots>i_k>0}
     \frac{1}{{i_1}^{m_1}}\ldots \frac{1}{{i_k}^{m_k}}.
\eq
This motivates the following definition of a special form of nested sums, called 
$Z$-sums \cite{Moch:2001zr,Weinzierl:2002hv,Weinzierl:2004bn,Moch:2005uc}:
\bq 
\label{definition}
  Z(n;m_1,...,m_k;x_1,...,x_k) & = & \sum\limits_{n\ge i_1>i_2>\ldots>i_k>0}
     \frac{x_1^{i_1}}{{i_1}^{m_1}}\ldots \frac{x_k^{i_k}}{{i_k}^{m_k}}.
\eq
$k$ is called the depth of the $Z$-sum and $w=m_1+...+m_k$ is called the weight.
If the sums go to infinity ($n=\infty$) the $Z$-sums are multiple polylogarithms:
\bq
\label{multipolylog}
Z(\infty;m_1,...,m_k;x_1,...,x_k) & = & \mbox{Li}_{m_1,...,m_k}(x_1,...,x_k).
\eq
For $x_1=...=x_k=1$ the definition reduces to the Euler-Zagier sums \cite{Euler,Zagier,Vermaseren:1998uu,Blumlein:1998if,Blumlein:2003gb}:
\bq
Z(n;m_1,...,m_k;1,...,1) & = & Z_{m_1,...,m_k}(n).
\eq
For $n=\infty$ and $x_1=...=x_k=1$ the sum is a multiple $\zeta$-value \cite{Borwein,Blumlein:2009}:
\bq
Z(\infty;m_1,...,m_k;1,...,1) & = & \zeta_{m_1,...,m_k}.
\eq
The usefulness of the $Z$-sums lies in the fact, that they interpolate between
multiple polylogarithms and Euler-Zagier sums.
The $Z$-sums form a quasi-shuffle algebra.
In this approach multiple polylogarithms appear through eq.~(\ref{multipolylog}).

Let us look as an example again at eq.~(\ref{basic_scalar_int}). Setting $D=4-2\eps$ we obtain:
\bq
\label{integralresult}
I & = &
 \int \frac{d^{4-2\eps}k_1}{i \pi^{2-\eps}}
 \frac{1}{(-k_1^2)}
 \frac{1}{(-k_2^2)}
 \frac{1}{(-k_3^2)}
 = 
 \left( - s_{123} \right)^{-1-\eps}
 \frac{\Gamma(-\eps)\Gamma(1-\eps)}{\Gamma(1-2\eps)}
 \sum\limits_{n=1}^\infty
 \frac{\Gamma(n+\eps)}
      {\Gamma(n+1)}
 \left(1-x\right)^{n-1},
 \nonumber \\
\eq
with $x=(-s_{12})/(-s_{123})$.
The simplest way to arrive at the sum representation is to use the following Feynman parametrisation:
\bq
 I
 & = & 
 \left( - s_{123} \right)^{-1-\eps}
 \Gamma(1+\eps)
 \int\limits_0^1 da \; 
 \int\limits_0^1 db \; b^{-\eps-1} (1-b)^{-\eps}
 \left[ 1- a \left(1-x\right) \right]^{-1-\eps}.
\eq
One then expands
$\left[ 1- a \left(1-x\right) \right]^{-1-\eps}$ according to
\bq
\left(1-z\right)^{-c} & = & 
 \frac{1}{\Gamma(c)} \sum\limits_{n=0}^\infty
 \frac{\Gamma(n+c)}{\Gamma(n+1)} z^n.
\eq
We continue with eq.~(\ref{integralresult}):
Expanding $\Gamma(n+\eps)$ according to eq.~(\ref{expansiongamma}) 
one obtains:
\bq
I
 & = & 
  \frac{\Gamma(-\eps)\Gamma(1-\eps)\Gamma(1+\eps)}{\Gamma(1-2\eps)}
 \frac{\left( - s_{123} \right)^{-1-\eps}}{1-x}
 \sum\limits_{n=1}^\infty
 \eps^{n-1}
 H_{\underbrace{1,...,1}_{n}}(1-x).
 \nonumber
\eq
In this special case all
harmonic polylogarithms can be expressed in terms of powers of the standard logarithm:
\bq
H_{\underbrace{1,...,1}_{n}}(1-x) & = & 
 \frac{(-1)^n}{n!} \left( \ln x \right)^n.
\eq
This particular example is very simple and one recovers the well-known
all-order result
\bq
  \frac{\Gamma(1-\eps)^2\Gamma(1+\eps)}{\Gamma(1-2\eps)}
 \frac{\left( - s_{123}^2 \right)^{-1-\eps}}{\eps^2}
 \frac{1-x^{-\eps}}{1-x},
\eq
which (for this simple example)
can also be obtained by direct integration. 
If we expand this result in $\eps$ we recover eq.~(\ref{final_result}).

An alternative approach to the computation of Feynman parameter integrals is based on
differential equations \cite{Kotikov:1990kg,Kotikov:1991pm,Remiddi:1997ny,Gehrmann:1999as,Gehrmann:2000zt,Gehrmann:2001ck,Argeri:2007up}.
To evaluate these integrals within this approach 
one first finds for each 
master integral a differential
equation, which this master integral has to satisfy.
The derivative is taken with respect to an external scale, or a
ratio of two scales.
An example for a one-loop four-point function is given by
\bq
\lefteqn{
\frac{\partial}{\partial s_{123}}
\begin{picture}(140,40)(-15,45)
\Vertex(50,20){2}
\Vertex(50,80){2}
\Vertex(20,50){2}
\Vertex(80,50){2}
\Line(0,50)(20,50)
\Line(20,50)(50,80)
\Line(50,20)(20,50)
\Line(50,80)(80,50)
\Line(80,50)(50,20)
\Line(50,80)(70,80)
\Line(80,50)(100,50)
\Line(50,20)(70,20)
\Text(75,80)[l]{\tiny $p_1$}
\Text(105,50)[l]{\tiny $p_2$}
\Text(75,20)[l]{\tiny $p_3$}
\end{picture}
= 
\frac{D-4}{2(s_{12}+s_{23}-s_{123})}
\begin{picture}(100,40)(-15,45)
\Vertex(50,20){2}
\Vertex(50,80){2}
\Vertex(20,50){2}
\Vertex(80,50){2}
\Line(0,50)(20,50)
\Line(20,50)(50,80)
\Line(50,20)(20,50)
\Line(50,80)(80,50)
\Line(80,50)(50,20)
\Line(50,80)(70,80)
\Line(80,50)(100,50)
\Line(50,20)(70,20)
\Text(75,80)[l]{\tiny $p_1$}
\Text(105,50)[l]{\tiny $p_2$}
\Text(75,20)[l]{\tiny $p_3$}
\end{picture}
} & &
\nonumber \\
& & \nonumber \\
 & &
 + \frac{2(D-3)}{(s_{123}-s_{12})(s_{123}-s_{12}-s_{23})}
\left[ 
\frac{1}{s_{123}}
\begin{picture}(110,40)(-5,45)
\Vertex(30,50){2}
\Vertex(70,50){2}
\Line(10,50)(30,50)
\CArc(50,50)(20,0,360)
\Line(70,50)(90,50)
\Text(80,55)[lb]{\tiny $p_{123}$}
\end{picture}
-
\frac{1}{s_{12}}
\begin{picture}(110,40)(-5,45)
\Vertex(30,50){2}
\Vertex(70,50){2}
\Line(10,50)(30,50)
\CArc(50,50)(20,0,360)
\Line(70,50)(90,50)
\Text(80,55)[lb]{\tiny $p_{12}$}
\end{picture}
\right]
\nonumber \\
 & &
 + \frac{2(D-3)}{(s_{123}-s_{23})(s_{123}-s_{12}-s_{23})}
\left[ 
\frac{1}{s_{123}}
\begin{picture}(110,40)(-5,45)
\Vertex(30,50){2}
\Vertex(70,50){2}
\Line(10,50)(30,50)
\CArc(50,50)(20,0,360)
\Line(70,50)(90,50)
\Text(80,55)[lb]{\tiny $p_{123}$}
\end{picture}
-
\frac{1}{s_{23}}
\begin{picture}(110,40)(-5,45)
\Vertex(30,50){2}
\Vertex(70,50){2}
\Line(10,50)(30,50)
\CArc(50,50)(20,0,360)
\Line(70,50)(90,50)
\Text(80,55)[lb]{\tiny $p_{23}$}
\end{picture}
\right].
\nonumber
\eq
The two-point functions on the r.h.s are simpler and can be considered to be known.
This equation is solved iteratively by an ansatz 
for the solution as a Laurent expression in $\eps$.
Each term in this Laurent series is a sum of terms, consisting of
basis functions times
some unknown (and to be determined) coefficients.
This ansatz is inserted into the differential equation and the unknown 
coefficients
are determined order by order from the differential equation.
The basis functions are taken as a subset of multiple polylogarithms.
In this approach the iterated integral representation of multiple polylogarithms is the most
convenient form. This is immediately clear from the
simple formula for the derivative as in eq.~(\ref{derivative}).

\section{Conclusions}
\label{sect:conclusions}

In these lectures I discussed Feynman integrals. After an introduction into the basic techniques,
the lectures focused on the computation of Feynman parameter integrals, with an
emphasis on the mathematical structures underlying these computations.
One encounters iterated structures as nested sums or iterated integrals, which form
a Hopf algebra with a shuffle or quasi-shuffle product.
Of particular importance are multiple polylogarithms. 
The algebraic properties of these functions are very rich: They form at the same time
a shuffle algebra as well as a quasi-shuffle algebra.
Based on these algebraic structures I discussed algorithms which evaluate Feynman integrals
to multiple polylogarithms.

\bibliography{/home/stefanw/notes/biblio}
\bibliographystyle{/home/stefanw/latex-style/h-physrev5}

\end{document}